\documentclass[aps,prd,twocolumn,superscriptaddress,nofootinbib,floatfix,longbibliography]{revtex4-2}

\usepackage{amsmath,amssymb}
\usepackage{bm}
\usepackage{graphicx}
\usepackage{xcolor}
\usepackage{booktabs}
\usepackage{placeins} 
\usepackage[colorlinks=true,linkcolor=blue!60!black,citecolor=blue!60!black,urlcolor=blue!60!black]{hyperref}

\newcommand{\Dbh}{\hat{\bar\nabla}}
\newcommand{\dd}{\mathrm{d}}

\newcommand{\sgb}{\sqrt{\bar g}}
\newcommand{\Tr}{\mathrm{Tr}}
\newcommand{\STr}{\mathrm{STr}}
\newcommand{\order}{\mathcal{O}}
\newcommand{\Nc}{N_{C}}
\newcommand{\Nr}{N_{R}}
\newcommand{\NHD}{N_{\rm HD}}
\newcommand{\fp}[1]{\mathrm{FP}_{#1}}

\newcommand{\figslot}[3][\columnwidth]{%
  \IfFileExists{figures/#2}{%
    \includegraphics[width=#1]{figures/#2}%
  }{%
    \fbox{\parbox[c][0.55#1][c]{0.92#1}{\centering\small
      \textbf{Figure placeholder}\\[2pt]\texttt{\detokenize{figures/#2}}\\[4pt]
      \parbox{0.85#1}{\centering\footnotesize #3}}}%
  }%
}

\DeclareRobustCommand{\fabrizio}[1]{{\color{green!50!black}#1}}

\begin{document}


\title{\texorpdfstring{The running of Weyl gauged gravity:\\
asymptotic freedom, its obstruction, and a way out}{The running of Weyl gauged gravity}}

\author{Diego Buccio}
\email{buccio@thphys.uni-heidelberg.de}
\affiliation{Institut f\"ur Theoretische Physik, Universit\"at Heidelberg, Philosophenweg 16, 69120 Heidelberg, Germany}

\author{Fabrizio Chicconi}
\email{fabrizio.chicconi@phd.unipi.it}
\affiliation{Universit\`a di Pisa and INFN -- Sezione di Pisa, Largo Bruno Pontecorvo 3, 56127 Pisa, Italy}

\author{Luca Parente}
\email{luca.parente@phd.unipi.it}
\affiliation{Universit\`a di Pisa and INFN -- Sezione di Pisa, Largo Bruno Pontecorvo 3, 56127 Pisa, Italy}

\author{Omar Zanusso}
\email{omar.zanusso@unipi.it}
\affiliation{Universit\`a di Pisa and INFN -- Sezione di Pisa, Largo Bruno Pontecorvo 3, 56127 Pisa, Italy}

\date{\today}

\begin{abstract}
We compute the one-loop beta functions of quadratic gravity in Weyl conformal geometry, the gauge theory of diffeomorphisms and dilatations with a dynamical Weyl gauge boson. The divergences are extracted diagrammatically in the background field method, with a gauge fixing procedure adapted to the semidirect product structure of the gauge group. The Weyl-tensor coupling and the Weyl gauge coupling are asymptotically free, and the running of the latter is special, being independent of the other couplings and of any two-derivative matter content. The obstruction to an ultraviolet completion comes from the square of a Weyl-covariant curvature scalar,
denoted $\hat R^2$:
the flow of the corresponding coupling has no fixed point in the physically interesting region and becomes trivial, the only real fixed points require a decoupled Weyl boson and a tachyonic scalar mode.
This also holds when including the Standard Model with any nonminimal Higgs coupling and, more generally, for any matter with two-derivative kinetic terms. Weyl-invariant four-derivative scalars with a sufficiently large coupling to $\hat R^2$ can remove the obstruction, producing a fixed point in the physical region that is ultraviolet attractive in all directions and has a nonvanishing Weyl gauge coupling.
\end{abstract}

\maketitle

\section{Introduction}
\label{sec:intro}

Scale symmetry is a recurring theme in the search for a fundamental description of gravity and particle physics. Observationally, the spectrum of primordial curvature perturbations measured in the cosmic microwave background is almost scale invariant, with scalar spectral index $n_s = 0.9649\pm 0.0042$~\cite{Planck:2018jri}; the small but nonzero deviation from $n_s=1$ indicates that scale symmetry must eventually be broken, for instance to end inflation after a finite number of $e$-folds. On the particle physics side, the Standard Model (SM) with vanishing Higgs mass parameter is classically scale invariant. Since global continuous symmetries are expected to be broken by quantum gravity effects~\cite{Kallosh:1995hi}, it is natural to consider a \emph{gauged} scale symmetry.

Gravitational actions quadratic in the curvatures are the natural gravitational counterpart of this program. Their couplings are dimensionless, so that the classical theory is scale invariant; they are also perturbatively renormalizable in four dimensions~\cite{Stelle:1976gc}, and the coupling of the Weyl-tensor squared term is asymptotically free~\cite{Fradkin:1981iu,Avramidi:1985ki,deBerredo-Peixoto:2004cjk,Codello:2006in,Salvio:2014soa,Salvio:2017qkx,Percacci:2017fkn,OhtaPercacci:2013,OhtaPercacci:2015}. The price for renormalizability is a massive spin-$2$ state with negative norm~\cite{Stelle:1976gc}, whose implications are still debated~\cite{Donoghue:2019fcb,Salvio:2018crh,Maldacena:2011mk}. It was also observed that the purely quadratic, scale invariant theory naturally recovers Einstein gravity in suitable backgrounds or phases~\cite{Alvarez-Gaume:2015rwa}.

A particularly economical way to combine these ingredients is provided by Weyl conformal geometry. Here the metric $g_{\mu\nu}$ and a vector field $\omega_\mu$, playing the role of the gauge boson of dilatations, form equivalence classes under local rescalings, and the most general action quadratic in the curvatures contains the terms $\hat R^2$ and $\hat F^2$ already considered by Weyl~\cite{Weyl:1918ib,Weyl:1919fi} ($\hat{R}$ is a Weyl covariant generalization of the scalar curvature and $\hat F_{\mu\nu}$ is the curvature $2$-form of $\omega_\mu$) and the square of the Weyl tensor~\cite{Ghilencea:2021lpa}. Originally, Weyl's theory was abandoned because of Einstein's objection about the path dependence of lengths (second clock effect), but in the modern, Weyl-gauge-covariant formulation, the geometry is metric and the objection can be circumvented~\cite{Ghilencea:2022lcl,Hobson:2020doi,Condeescu:2023izl,Ghilencea:2026zer}. A series of recent works by Ghilencea and collaborators~\cite{Ghilencea:2018dqd,Ghilencea:2019jux,Ghilencea:2021lpa,Ghilencea:2023sti,Ghilencea:2024usf,Ghilencea:2026zer,Ghilencea:2026jgp} has shown that this theory, hereafter called Weyl quadratic gravity, has several remarkable properties:
(i) it is the gauge theory of the Weyl group (Poincar\'e and dilatations) with a physical gauge boson~\cite{Charap:1974cr,Condeescu:2024cjh,Viacava:2026kbz};
(ii) the dilatation symmetry is broken spontaneously by a \emph{geometric} St\"uckelberg mechanism, in which $\omega_\mu$ absorbs the would-be Goldstone mode contained in the $\hat R^2$ term and becomes a massive Proca field, while the Einstein--Hilbert action and a positive cosmological constant emerge in the broken phase, so that the Planck scale, the cosmological constant and the Weyl boson mass all have a common geometric origin~\cite{Ghilencea:2018dqd,Ghilencea:2019jux};
(iii) the SM embeds naturally, with only the Higgs field coupling directly to $\omega_\mu$; the Higgs mass may then be protected by the gauged scale symmetry, and electroweak symmetry breaking can be induced gravitationally~\cite{Ghilencea:2021lpa,Ghilencea:2023wwf};
(iv) since the Weyl-covariant scalar curvature transforms covariantly, the theory admits a geometric, Weyl-invariant regularization and has been argued to be free of Weyl anomalies~\cite{Ghilencea:2023sti,Ghilencea:2024usf};
(v) it leads to Starobinsky-like inflation~\cite{Ghilencea:2019rqj} (see also Refs.~\cite{BeltranJimenez:2015VD,GialamasTamvakis:2024} for related constructions in generalized Weyl and Einstein--Cartan geometries) and to specific signatures such as vector polarizations of gravitational waves~\cite{Ghilencea:2026jgp}.

The natural question, raised explicitly in Ref.~\cite{Ghilencea:2024usf}, is whether this theory can be a consistent quantum field theory up to arbitrarily high energies, i.e., whether its renormalization group (RG) flow admits a UV completion. In this paper we answer this question at one loop. We compute the complete one-loop gauge-independent beta functions of Weyl quadratic gravity, in pure gravity and in the presence of matter, and we analyze the resulting flow. Previous studies addressed related questions in Weyl-invariant theories linear in the curvature~\cite{Percacci:2011uf,Codello:2012sn,Pagani:2013fca}. Here we use a diagrammatic approach: we work in the background field method~\cite{DeWitt:1967ub,Abbott:1981ke}, we compute the divergent parts of the two-point functions of the background metric and of the background Weyl vector from tadpole and bubble diagrams, and we isolate the beta functions by projecting onto spin components~\cite{Percacci:2017fkn}. Two technical aspects require care. First, the gauge group is the semidirect product of diffeomorphisms and Weyl transformations; a background-covariant gauge-fixing is obtained only after an improvement of the diffeomorphism generators~\cite{Jackiw:1978ar,Daum:2009dn,Sauro:2022chz}. Second, the higher-derivative gauge-fixing of diffeomorphisms requires the inclusion of the third (Nielsen--Kallosh) ghost~\cite{Nielsen:1978mp,Kallosh:1978de,Fradkin:1981iu,Barth:1983hb,Ohta:2016jvw}.

Our main results are the following. Both the Weyl-tensor coupling $\eta$ and the Weyl gauge coupling $\alpha$ are asymptotically free. The running of $\alpha$ does not depend on the other couplings nor on the ordinary matter content, i.e., fields with two-derivative kinetic terms, a feature that is unusual for an Abelian gauge symmetry and that we trace back to the fact that the fields charged under dilatations which couple to the transverse Weyl boson belong to the gravitational sector. The obstruction to a UV completion comes instead from the $\hat R^2$ sector. Expressing the flow in terms of the ratios of couplings, we find four fixed points: two complex conjugate ones, and two real ones which require $\alpha=0$ and $\xi^2<0$, i.e., a decoupled Weyl boson and a tachyonic scalar sector. Inside the physical region the coupling $\xi$ leaves the perturbative regime at a finite scale. We prove that no matter content with non-negative contribution to the $\hat R^2$ divergence, including the SM for any value of the Higgs nonminimal coupling, can modify this picture. Finally, we show that Weyl-invariant, dimensionless four-derivative scalars~\cite{Paneitz:2008afy,Fradkin:1985am,Boyle:2021jaz,Buccio:2024omv} with a positive nonminimal coupling to $\hat R^2$ provide a possible missing ingredient for UV completeness: the complex fixed points become real and one of them moves into the physical region, where it is UV attractive in all directions and has a nonvanishing Weyl gauge coupling.

The paper is organized as follows. In Sec.~\ref{sec:geometry} we review Weyl conformal geometry, the action and the physical meaning of the couplings. Section~\ref{sec:quantization} describes the background field quantization, gauge fixing procedure and ghosts. In Sec.~\ref{sec:method} we explain the spin projections and the diagrammatic extraction of the divergences. The beta functions are presented in Sec.~\ref{sec:betas} and the RG flow of pure gravity is analyzed in Sec.~\ref{sec:flow}. The effect of SM matter is discussed in Sec.~\ref{sec:SM}, and the higher-derivative scalars in Sec.~\ref{sec:HD}. We conclude in Sec.~\ref{sec:conclusions}. Technical details are collected in the Appendices.

\section{Weyl conformal geometry and the action}
\label{sec:geometry}

\subsection{Weyl gauge symmetry and Weyl-covariant geometry}

We follow the Weyl-gauge-covariant formulation of Weyl conformal geometry of Refs.~\cite{Ghilencea:2023sti,Condeescu:2023izl,Ghilencea:2026zer,Ghilencea:2026jgp}, and work in Euclidean signature in four dimensions. Weyl geometry is defined by equivalence classes of a metric $g_{\mu\nu}$ and of a vector field $\omega_\mu$, related by local Weyl gauge transformations
\begin{equation}
g_{\mu\nu}\to \Sigma^2 g_{\mu\nu}\,,\qquad
\omega_\mu \to \omega_\mu-\partial_\mu\ln\Sigma\,,\qquad
\Phi\to\Sigma^{q_\Phi}\Phi\,,
\label{eq:weyltransf}
\end{equation}
with $\Sigma(x)=e^{\sigma(x)}>0$. Here $\Phi$ denotes a generic tensor field of Weyl charge (or weight) $q_\Phi$; the metric has $q_g=2$, its inverse $-2$, and $\sqrt g$ has charge $4$. Infinitesimally, $\delta^{\rm(W)}_\sigma g_{\mu\nu}=2\sigma g_{\mu\nu}$, $\delta^{\rm(W)}_\sigma\omega_\mu=-\partial_\mu\sigma$ and $\delta^{\rm(W)}_\sigma\Phi=q_\Phi\sigma\Phi$. Since the group of dilatations is Abelian, the normalization of the charges is a convention; with the choice $q_g=2$ the gauge coupling does not appear in Eq.~\eqref{eq:weyltransf} and will appear only in front of the kinetic term of $\omega_\mu$.\footnote{In Ref.~\cite{Ghilencea:2021lpa} the metric has charge $1$ and the coupling, there denoted $\alpha$, sits in the covariant derivative. The two descriptions are related by a rescaling of $\omega_\mu$; physical quantities such as the Weyl boson mass in Eq.~\eqref{eq:scales} coincide.}

The Weyl connection $\tilde\Gamma$ is the symmetric torsion-free connection satisfying the non-metricity condition $\tilde\nabla_\lambda g_{\mu\nu}=-2\,\omega_\lambda g_{\mu\nu}$,
\begin{equation}
\tilde\Gamma^\rho{}_{\mu\nu}=\Gamma^\rho{}_{\mu\nu}+\delta^\rho_\mu\,\omega_\nu+\delta^\rho_\nu\,\omega_\mu-g_{\mu\nu}\,\omega^\rho\,,
\label{eq:weylconnection}
\end{equation}
where $\Gamma$ is the Levi-Civita connection. $\tilde\Gamma$ is invariant under Eq.~\eqref{eq:weyltransf}, but the derivative $\tilde\nabla$ of a charged tensor does not transform covariantly. The Weyl-covariant derivative is defined by covariantizing with respect to the charge of the field on which it acts~\cite{Ghilencea:2023sti,Condeescu:2023izl},
\begin{equation}
\hat\nabla_\mu T=\left(\tilde\nabla_\mu+q_T\,\omega_\mu\right)T\,,\qquad
\hat\nabla'_\mu T'=\Sigma^{q_T}\hat\nabla_\mu T\,.
\label{eq:covder}
\end{equation}
In particular $\hat\nabla_\lambda g_{\mu\nu}=0$: in the Weyl-covariant formulation the geometry is \emph{metric}, the length $g_{\mu\nu}V^\mu V^\nu$ of a vector of charge $-1$ is preserved by Weyl-covariant parallel transport, and no second clock effect arises~\cite{Ghilencea:2022lcl,Hobson:2020doi,Ghilencea:2026zer}. For example, on a scalar of charge (weight) $q$ one simply has $\hat\nabla_\mu\phi=(\partial_\mu+q\,\omega_\mu)\phi$.\footnote{%
The coupling of $\omega_\mu$ with fields with half-integer spins is slightly more complicate in that they require a spin-connection for the local Lorentz group, but the connection $\tilde \Gamma$ would require coupling to the full $GL(4)$ group
of the tangent space, which includes its symmetric generators \cite{Viacava:2026kbz}. However it is always possible to trade $\omega_\mu$ for a torsion vector that, instead, couples naturally to half-integer spins using a lightcone projective symmetry (referred to as torsion gauging in Ref.~\cite{Sauro:2022hoh}; see also Ref.~\cite{IosifidisKoivisto:2018} for the different notions of scale transformation in metric-affine geometry).} In practice, for our purposes this does not play a role: the Dirac operator is Weyl covariant and the Dirac action is Weyl invariant in any dimension
without the need for a direct coupling to $\omega_\mu$.

The curvature follows from the commutator of Weyl-covariant derivatives. On a vector of charge $q_V$,
\begin{equation}
\begin{aligned}
[\hat\nabla_\mu,\hat\nabla_\nu]V^\rho&=\tilde R^\rho{}_{\sigma\mu\nu}V^\sigma+q_V\,\hat F_{\mu\nu}V^\rho\\
&=\hat R^\rho{}_{\sigma\mu\nu}V^\sigma+(q_V+1)\,\hat F_{\mu\nu}V^\rho\,,
\end{aligned}
\label{eq:commutator}
\end{equation}
where $\tilde R^\rho{}_{\sigma\mu\nu}$ is the Riemann tensor of $\tilde\Gamma$ and $\hat F_{\mu\nu}=\partial_\mu\omega_\nu-\partial_\nu\omega_\mu$. Since $\tilde\Gamma$ is not metric, $\tilde R_{\rho\sigma\mu\nu}=\hat R_{\rho\sigma\mu\nu}+g_{\rho\sigma}\hat F_{\mu\nu}$ contains, besides the part $\hat R_{\rho\sigma\mu\nu}$ antisymmetric in $\rho\sigma$, the homothetic curvature $\tilde R^\rho{}_{\rho\mu\nu}=4\hat F_{\mu\nu}$. The Riemann tensor of Weyl geometry $\hat R^\rho{}_{\sigma\mu\nu}$ is thus given by the commutator on vectors of charge $-1$, i.e., with Weyl-neutral frame components $V^a=e^a{}_\rho V^\rho$~\cite{Ghilencea:2026jgp}, while the dilatation part of the curvature is carried entirely by $\hat F_{\mu\nu}$. In terms of Riemannian quantities one finds~\cite{Ghilencea:2026jgp}
\begin{align}
\hat R_{\mu\nu}&=R_{\mu\nu}-2\nabla_\nu\omega_\mu-g_{\mu\nu}\nabla_\alpha\omega^\alpha\nonumber\\
&\quad+2\left(\omega_\mu\omega_\nu-g_{\mu\nu}\,\omega_\alpha\omega^\alpha\right),\label{eq:Ricci}\\
\hat R&=R-6\,\nabla_\mu\omega^\mu-6\,\omega_\mu\omega^\mu\,,\label{eq:Rhat}
\end{align}
where $\nabla$ is the Levi-Civita covariant derivative and $\hat R_{\mu\nu}=\hat R^\sigma{}_{\mu\sigma\nu}$, whose antisymmetric part is $\hat F_{\mu\nu}$. Under Weyl transformations $\hat R^\rho{}_{\sigma\mu\nu}$ and $\hat R_{\mu\nu}$ are invariant, while $\hat R\to\Sigma^{-2}\hat R$. Moreover, the Weyl tensor of $\hat R^\rho{}_{\sigma\mu\nu}$ coincides with the Riemannian one, $\hat C^\rho{}_{\sigma\mu\nu}=C^\rho{}_{\sigma\mu\nu}$~\cite{Condeescu:2023izl,Ghilencea:2026jgp}. As a consequence, in $d=4$, the densities
\begin{equation}
\sqrt g\,C_{\mu\nu\rho\sigma}C^{\mu\nu\rho\sigma}\,,\qquad
\sqrt g\,\hat R^2\,,\qquad
\sqrt g\,\hat F_{\mu\nu}\hat F^{\mu\nu}
\end{equation}
are separately Weyl invariant. The Euler density $\hat G$ of Weyl geometry is a total derivative in $d=4$ and thus will not be included in this work; in $d=4$, the Gauss--Bonnet-type combinations built from the different contractions of the Weyl curvature differ from it only by $\hat F^2$ terms, up to total derivatives~\cite{BeltranJimenez:2014GB}.\footnote{If one uses instead $\tilde R^\rho{}_{\sigma\mu\nu}$, as in Weyl's original formulation and in Ref.~\cite{Chicconi:2024thesis}, the homothetic part contributes to the Weyl tensor, $\tilde C^2=C^2+6\hat F^2$. This only reshuffles part of the $\hat F^2$ term into the Weyl-squared term and amounts to a linear redefinition of the couplings: the coefficient of the full kinetic term of $\omega_\mu$, which is what enters physical quantities, is unaffected. We use the basis $\hat C=C$, in which $\alpha$ multiplies the full kinetic term.}

The Weyl-covariant formulation also clarifies the structure of the gauge group. Under a diffeomorphism generated by $\epsilon^\mu$, the fields transform as $\delta^{\rm(D)}_\epsilon g_{\mu\nu}=\mathcal L_\epsilon g_{\mu\nu}$ and $\delta^{\rm(D)}_\epsilon\omega_\mu=\mathcal L_\epsilon\omega_\mu$. The commutator of a diffeomorphism and a Weyl transformation is again a Weyl transformation, $[\delta^{\rm(D)}_\epsilon,\delta^{\rm(W)}_\sigma]=\delta^{\rm(W)}_{\mathcal L_\epsilon\sigma}$, while Weyl transformations commute among themselves. The gauge group is therefore the semidirect product
\begin{equation}
\mathcal G=\mathrm{Diff}(M)\ltimes\mathrm{Weyl}(M)\,,
\end{equation}
with the Weyl transformations forming an invariant subgroup~\cite{Chicconi:2024thesis}. The Lie derivative is not covariant with respect to Weyl transformations. A Weyl-covariant version of diffeomorphisms is obtained by combining each diffeomorphism with a field-dependent Weyl transformation of parameter $\sigma=\epsilon^\mu\omega_\mu$~\cite{Jackiw:1978ar,Iorio:1996ad,Daum:2009dn,Sauro:2022chz},
\begin{equation}
\tilde\delta^{\rm(D)}_\epsilon\equiv\delta^{\rm(D)}_\epsilon+\delta^{\rm(W)}_{\sigma=\epsilon\cdot\omega}\,,
\label{eq:improved}
\end{equation}
which acts as
\begin{equation}
\begin{gathered}
\tilde\delta^{\rm(D)}_\epsilon g_{\mu\nu}=\hat\nabla_\mu\epsilon_\nu+\hat\nabla_\nu\epsilon_\mu\,,\\
\tilde\delta^{\rm(D)}_\epsilon\omega_\mu=\epsilon^\nu\hat F_{\nu\mu}\,,\qquad
\tilde\delta^{\rm(D)}_\epsilon\phi=\epsilon^\mu\hat\nabla_\mu\phi\,,
\end{gathered}
\label{eq:improvedaction}
\end{equation}
where $\epsilon_\mu=g_{\mu\nu}\epsilon^\nu$ carries charge 2. In the improved basis the Weyl transformations commute with the improved diffeomorphisms, and the improved diffeomorphisms close up to a field-dependent Weyl transformation with parameter proportional to $\hat F_{\mu\nu}\epsilon_1^\mu\epsilon_2^\nu$~\cite{Chicconi:2024thesis}. This disentangles the two symmetries and will allow us to define gauge conditions and ghost actions that preserve the full background gauge symmetry (Sec.~\ref{sec:quantization}).

\subsection{Action} 
The most general local action built from Weyl-invariant densities without dimensionful couplings is quadratic in the curvatures~\cite{Weyl:1918ib,Weyl:1919fi,Ghilencea:2021lpa}. In Euclidean signature, which we use throughout, it reads
\begin{equation}
S=\int\dd^4x\sqrt g\left[\frac{C^2}{\eta^2}-\frac{\hat R^2}{4!\,\xi^2}+\frac{\hat F^2}{4\alpha^2}\right],
\label{eq:action}
\end{equation}
with $C^2\equiv C_{\mu\nu\rho\sigma}C^{\mu\nu\rho\sigma}$, $\hat F^2\equiv\hat F_{\mu\nu}\hat F^{\mu\nu}$ and dimensionless couplings $\eta$, $\xi$ and $\alpha$.\footnote{For the Lorentzian form of the action we refer to Refs.~\cite{Ghilencea:2021lpa,Ghilencea:2024usf,Ghilencea:2026zer,Ghilencea:2026jgp}. Equation~\eqref{eq:action} follows from it by Wick rotation, with the couplings $\eta$, $\xi$ and $\alpha$ normalized as in Refs.~\cite{Ghilencea:2026zer,Ghilencea:2026jgp}; for the normalization of Ref.~\cite{Ghilencea:2021lpa} see Appendix~\ref{app:conventions}.}
The three terms are separately invariant under diffeomorphisms and under Eq.~\eqref{eq:weyltransf}. The coupling $\eta$ controls the conformal (Weyl-tensor) sector, $\xi$ the $\hat R^2$ sector, which contains the scalar mode of the metric and its coupling to $\omega_\mu$ through Eq.~\eqref{eq:Rhat}, and $\alpha$ is the Weyl gauge coupling. Expanding Eq.~\eqref{eq:Rhat}, the $\hat R^2$ term contains $R^2$, the mixing $-12R\,\nabla_\mu\omega^\mu$, the longitudinal kinetic term $36(\nabla_\mu\omega^\mu)^2$ and interaction terms; the transverse kinetic term of $\omega_\mu$ only comes from $\hat F^2$. Note that the $\hat R^2$ term has negative coefficient for real $\xi$; this is the sign required by the absence of tachyons in the scalar sector, as discussed in Sec.~\ref{sec:physregion}.

\subsection{Broken phase and physical meaning of the couplings}
\label{sec:brokenphase}
The physical meaning of the couplings is more transparent in the spontaneously broken phase~\cite{Ghilencea:2018dqd,Ghilencea:2019jux,Ghilencea:2021lpa}. In Eq.~\eqref{eq:action} one replaces
\begin{equation}
\hat R^2\to2\phi^2\hat R-\phi^4\,,
\label{eq:linearization}
\end{equation}
with a real auxiliary scalar $\phi$ of charge $-1$; the two forms are classically equivalent on the solution $\phi^2=\hat R>0$ (we use the curvature convention in which the four-sphere has $R>0$).\footnote{Equation~\eqref{eq:linearization} is a special case of the general linearization $b\hat R^2\to\Phi\hat R-\Phi^2/(4b)$, with $\Phi=2b\hat R$ of mass dimension two, which describes both de Sitter and anti-de Sitter backgrounds~\cite{Alvarez-Gaume:2015rwa}.} This choice has two advantages: $\phi$ is a scalar of canonical dimension, and $\ln(\phi)$ shifts under Weyl transformations, playing the role of the would-be Goldstone boson (the dilaton) of gauged dilatations; moreover, $\phi^2\ge0$ selects the branch of positive curvature, i.e., the de Sitter vacuum in Lorentzian signature.

Fixing the gauge $\phi\to\langle\phi\rangle$ and using Eq.~\eqref{eq:Rhat}, up to a total derivative one obtains the Euclidean Einstein--Proca action with a positive cosmological constant,
\begin{multline}
S\to\int\dd^4x\sqrt{g}\Big[-\frac{M_p^2}{2}R+\Lambda M_p^2+\frac{\hat F^2}{4\alpha^2}\\
+3M_p^2\,\omega_\mu\omega^\mu+\frac{C^2}{\eta^2}\Big],
\end{multline}
in which the Weyl boson has become massive by absorbing the dilaton (St\"uckelberg mechanism). All scales are generated by the single vacuum expectation value $\langle\phi^2\rangle$, which is of geometric origin~\cite{Ghilencea:2018dqd,Ghilencea:2019jux,Ghilencea:2026jgp}:
\begin{equation}
\begin{aligned}
M_p^2&=\frac{\langle\phi^2\rangle}{6\xi^2}\,,&\qquad
\Lambda&=\frac{\langle\phi^2\rangle}{4}=\frac{3}{2}\xi^2M_p^2\,,\\
m_\omega^2&=6\,\alpha^2M_p^2\,,&\qquad
m_2^2&=\frac{\eta^2M_p^2}{4}\,.
\end{aligned}
\label{eq:scales}
\end{equation}
Here $M_p$ is the (reduced) Planck mass, $\Lambda$ the cosmological constant, $m_\omega$ the mass of the Weyl boson and $m_2$ the mass of the spin-$2$ state with negative norm associated with the Weyl-squared term~\cite{Stelle:1976gc,Alvarez-Gaume:2015rwa}. Below $m_\omega$ the Weyl boson decouples, $\tilde\Gamma\to\Gamma$, and Weyl geometry becomes Riemannian: Einstein gravity with $\Lambda>0$ is the low-energy phase of the theory~\cite{Ghilencea:2019jux,Ghilencea:2026zer}. The ratios of couplings, that will play a central role in the RG analysis below, acquire therefore a direct physical meaning,
\begin{equation}
\frac{\alpha^2}{\eta^2}=\frac{m_\omega^2}{24\,m_2^2}\,,\qquad
\frac{\xi^2}{\eta^2}=\frac{\Lambda}{6\,m_2^2}\,.
\label{eq:ratios_masses}
\end{equation}

\subsection{The physical region of coupling space}

\label{sec:physregion}
Equation~\eqref{eq:scales} determines which region of coupling space corresponds to a physically acceptable theory in the broken phase. We require:
\begin{itemize}
\item $\xi^2>0$. This guarantees $M_p^2>0$, i.e., the correct sign of the induced Einstein--Hilbert term, and $\Lambda>0$. For $\xi^2<0$ the scalar mode contained in $\hat R^2$ would be tachyonic. In the language of quadratic gravity this is the familiar non-tachyonic condition on the $R^2$ coupling.
\item $\alpha^2>0$. The Weyl boson has a kinetic term of the correct sign and, in the broken phase, a positive mass squared $m_\omega^2$.
\item $\eta^2>0$. The Weyl-squared term has the sign that makes the massive spin-$2$ state non-tachyonic. The state still has negative norm for any $\eta^2>0$; this is the well-known unitarity problem of quadratic gravity, which is not affected by the RG flow and which we do not address here~\cite{Stelle:1976gc,Donoghue:2019fcb,Salvio:2018crh,Maldacena:2011mk}.
\end{itemize}
In addition, the couplings should be small for perturbation theory to be reliable. Phenomenologically, Refs.~\cite{Ghilencea:2021lpa,Ghilencea:2024usf,Ghilencea:2026jgp} consider $\xi,\eta,\alpha<1$, with $\xi\ll\eta$ in the infrared, since $\xi^2\sim\Lambda/M_p^2$. We call \emph{physical region} the set of real couplings $\eta^2,\xi^2,\alpha^2>0$; the hierarchy $\xi<\eta$ will be indicated separately.

\subsection{Matter}
\label{sec:matter}

The Standard Model is embedded in Weyl geometry by promoting all derivatives to Weyl-covariant ones~\cite{Ghilencea:2021lpa,Ghilencea:2023wwf,Ghilencea:2026zer}. Gauge fields have charge zero and their action does not depend on $\omega_\mu$, because the connection~\eqref{eq:weylconnection} is symmetric.\footnote{%
Technically, a new coupling of the form $B_{\mu\nu}\hat F^{\mu\nu}$, for $B_{\mu\nu}$ the $2$-form curvature of an Abelian gauge theory such as SM's $U(1)_Y$, is allowed and would represent a nontrivial mixing with the vector $\omega_\mu$ with interesting phenomenology if interactions are considered \cite{Ghilencea:2021lpa}. In this paper we neglect the effect of SM's Yukawa interactions for simplicity.
} For fermions of charge $-3/2$ the dependence on $\omega_\mu$ cancels between the covariant derivative and the Weyl spin connection~\cite{Ghilencea:2021lpa}; in practice, the Dirac operator is Weyl-covariant. The only SM field coupled to $\omega_\mu$ is the Higgs doublet $H$, of charge $-1$, with Euclidean action
\begin{equation}
S_H=\int\dd^4x\sqrt g\left[|\hat\nabla_\mu H|^2-\frac{\xi_h}{6}|H|^2\hat R+\lambda_H|H|^4\right],
\label{eq:higgs}
\end{equation}
where $\hat\nabla_\mu H=(D_\mu-\omega_\mu)H$ and $D_\mu$ is the SM covariant derivative. Using Eq.~\eqref{eq:Rhat} and integrating by parts, all couplings to $\omega_\mu$ are proportional to $1+\xi_h$,
\begin{equation}
S_H\supset(1+\xi_h)\int\dd^4x\sqrt g\left[\omega_\mu\omega^\mu|H|^2-\omega^\mu\partial_\mu|H|^2\right],
\label{eq:higgsomega}
\end{equation}
so that for $\xi_h=-1$ the Higgs is conformally coupled and $\omega_\mu$ decouples up to a total derivative. In this sign convention, $\xi_h>0$ increases the Planck mass in the broken phase~\cite{Ghilencea:2021lpa}. Two features of Eq.~\eqref{eq:higgsomega} will be important: the cubic vertex couples $\omega_\mu$ to the gradient of $|H|^2$ and therefore only to the longitudinal component of $\omega_\mu$, while the quartic vertex produces a momentum-independent tadpole, which cannot contribute to terms with two derivatives of $\omega_\mu$. As a consequence, SM matter fields do not contribute to the renormalization of the transverse part of the $\omega_\mu$ propagator (see Sec.~\ref{sec:betas}).

At one loop and for vanishing matter backgrounds, free matter fields contribute to the gravitational divergences only through two numbers~\cite{Vassilevich:2003xt,Percacci:2017fkn}: their weight $\Nc$ in the Weyl-squared divergence and their weight $\Nr$ in the $\hat R^2$ divergence. Measured in units of a minimally coupled real scalar,
\begin{equation}
\Nc=N_S+6N_F+12N_V\,,\qquad
\Nr=\sum_{i=1}^{N_S}(1+\xi_i)^2\,,
\label{eq:NcNr}
\end{equation}
for $N_S$ real scalars with nonminimal couplings $\xi_i$ normalized as in Eq.~\eqref{eq:higgs}, $N_F$ Dirac fermions and $N_V$ gauge vectors. Fermions and gauge vectors, being conformally invariant, do not contribute to $\Nr$. For the SM with three right-handed neutrinos, $N_S=4$, $N_F=24$, $N_V=12$, so that
\begin{equation}
\Nc^{\rm SM}=292\,,\qquad \Nr^{\rm SM}=4(1+\xi_h)^2\,.
\end{equation}
Higher-derivative scalars are introduced in Sec.~\ref{sec:HD}.

\section{Quantization}
\label{sec:quantization}

\subsection{Background field method}

We split the fields into backgrounds and quantum fluctuations~\cite{DeWitt:1967ub,Abbott:1981ke},
\begin{equation}\label{eq:split}
g_{\mu\nu}=\bar g_{\mu\nu}+h_{\mu\nu}\,,\qquad \omega_\mu=\bar\omega_\mu+v_\mu\,.
\end{equation}
The gauge transformations of the total fields can be realized in two ways. In \emph{quantum} transformations the backgrounds are inert and the whole variation is assigned to the fluctuations; in \emph{background} transformations the backgrounds transform as in Eqs.~\eqref{eq:weyltransf} and~\eqref{eq:improvedaction}, while $h_{\mu\nu}$ and $v_\mu$ transform as tensors of charge $2$ and $0$, respectively. The gauge is fixed for the quantum transformations only as it should, but in a way that preserves the background transformations. 

The one-loop effective action is then a background-gauge-invariant functional, and its divergent part is a linear combination of the invariants of Eq.~\eqref{eq:action} evaluated in the background.\footnote{%
Strictly speaking, the effective action is a functional of both background and quantum fields. Assuming that the Ward identity of the arbitrary split introduced in Eq.~\eqref{eq:split}
is non-anomalous, which is true in dimensional regularization, the effective action depends only on the sum of background and quantum fields. Therefore, to recover the full effective action, it is sufficient to evaluate it in the limit of vanishing quantum fluctuations and background covariance of the final result implies full quantum covariance.
} Because of the semidirect product structure of $\mathcal G$, background invariance benefits from the use of the improved diffeomorphisms~\eqref{eq:improved}: with the standard Lie derivative, the Faddeev--Popov operator of the Weyl gauge condition would contain terms such as $\hat\nabla^\mu(\partial_\alpha\bar\omega_\mu\,c^\alpha+\bar\omega_\alpha\partial_\mu c^\alpha)$, which are not Weyl covariant~\cite{Chicconi:2024thesis}.

\subsection{Gauge fixing}

For diffeomorphisms we use the background-covariant condition
\begin{equation}
\chi_\mu=\Dbh^\nu h_{\nu\mu}+b\,\Dbh_\mu h\,,\qquad h=\bar g^{\mu\nu}h_{\mu\nu}\,,
\label{eq:chidiff}
\end{equation}
where $\Dbh$ is the Weyl-covariant derivative built with $\bar g_{\mu\nu}$ and $\bar\omega_\mu$. Since $h_{\mu\nu}$ has charge $2$, $\chi_\mu$ transforms covariantly under background Weyl transformations. Since the theory contains four derivatives, it is convenient to choose a gauge-fixing term that contains two additional derivatives,
\begin{equation}
S_{\rm gf}^{\rm D}=\frac{1}{2a}\int\dd^4x\sgb\;\chi_\mu Y^{\mu\nu}\chi_\nu\,,
\label{eq:gfdiff}
\end{equation}
\begin{equation}
Y^{\mu\nu}=\bar g^{\mu\nu}\,\Dbh^2+p_1\,\Dbh^\mu\Dbh^\nu-p_2\,\Dbh^\nu\Dbh^\mu\,.
\label{eq:Y}
\end{equation}
For the Weyl symmetry we use the background Lorenz condition
\begin{equation}
\chi=\Dbh^\mu v_\mu\,,\qquad
S_{\rm gf}^{\rm W}=\frac{1}{2\zeta}\int\dd^4x\sgb\;\chi^2\,,
\label{eq:gfweyl}
\end{equation}
with gauge parameter $\zeta$.

\subsection{Ghosts and effective action}

The ghost action is obtained in the BRST formalism~\cite{Ohta:2016jvw,Buchbinder:1992gdx}. We introduce diffeomorphism ghosts $c^\mu,\bar c_\mu$ and Weyl ghosts $\vartheta,\bar\vartheta$. With the improved generators~\eqref{eq:improvedaction}, the variations of the gauge conditions at lowest order in the fluctuations are
\begin{align}
\delta\chi_\mu&=\Delta_{\mu\nu}c^\nu+2(1+4b)\,\partial_\mu\vartheta\,,\\
\delta\chi&=-\Dbh^2\vartheta-\Dbh^\mu\big(\bar F_{\mu\nu}c^\nu\big)\,,
\end{align}
with $\Delta_{\mu\nu}=\bar g_{\mu\nu}\Dbh^2+(1+2b)\Dbh_\mu\Dbh_\nu+\hat{\bar R}_{\mu\nu}$, up to ordering of the derivatives. The ghost sector is therefore a $2\times2$ system with a mixing between the two symmetries. The term $2(1+4b)\,\bar c^\mu\partial_\mu\vartheta$ comes from the Weyl variation $\delta h_{\mu\nu}=2\vartheta\bar g_{\mu\nu}$ of the trace part of $\chi_\mu$, while the term $\bar\vartheta\,\Dbh^\mu(\bar F_{\mu\nu}c^\nu)$ is the improved diffeomorphism of the Weyl fluctuation. On a flat background with $\bar\omega_\mu=0$ the ghost kinetic matrix is triangular, so that a single mixed propagator $\langle c\,\bar\vartheta\rangle$ is generated, while for $\bar\omega_\mu\neq0$ the mixing also produces interaction vertices with the background Weyl vector.

The higher-derivative gauge-fixing~\eqref{eq:gfdiff} requires one more ingredient. Since $Y^{\mu\nu}$ is a differential operator that depends on the background, the Gaussian average over the gauge conditions brings a factor $(\det Y)^{1/2}$ into the functional measure, which, unlike in two-derivative gauges, is not a constant and contributes to the effective action~\cite{Nielsen:1978mp,Kallosh:1978de,Fradkin:1981iu,Barth:1983hb}. In the BRST formulation this factor arises automatically from the Nakanishi--Lautrup field, which is dynamical in the higher-derivative case~\cite{Ohta:2016jvw}. It is customarily attributed to a vector field $z_\mu$ with kinetic operator $Y$, the so-called third ghost, which contributes $-\frac12\Tr\ln Y$ to the effective action, i.e., as a real bosonic vector but with opposite sign.

The one-loop effective action reads
\begin{equation}
\Gamma^{(1)}=\frac12\,\STr\ln\mathcal H^{(h,v,\varphi)}-\Tr\ln\Delta^{\rm gh}-\frac12\,\Tr\ln Y\,,
\label{eq:Gamma1}
\end{equation}
where $\mathcal H^{(h,v,\varphi)}$ is the Hessian of $S+S^{\rm D}_{\rm gf}+S^{\rm W}_{\rm gf}$ with respect to the graviton, the Weyl fluctuation and the matter fields $\varphi$ (the supertrace accounting for fermions), $\Delta^{\rm gh}$ is the ghost operator including the mixing between diffeomorphism and Weyl ghosts, and the last term is the third ghost. The third ghost is essential: without it the known results of quadratic gravity are not recovered.

A concrete computational difficulty is that the Hessian operator ${\cal H}$ is not diagonal and has entries of different rank. Concentrating on the gravitational sector, in momentum space the fluctuations $h_{\mu\nu}$ have a $p^4$ higher-derivative propagator, $v_\mu$ have a standard $p^2$ propagator,
and the off-diagonal entries a $p^3$. Methods to compute $\Gamma^{(1)}$ covariantly with some nonminimal operators are known, see, for example, Ref.~\cite{Sauro:2025sbt} and references therein, but generally computationally intensive.
We discuss the strategy with which we circumvented this difficulty in the next section.

\subsection{Choice of gauge}

The only gauge parameter that we fix is $\zeta$, for which we take the Landau limit $\zeta\to0$. The parameters $a$, $b$, $p_1$ and $p_2$ of the diffeomorphism gauge-fixing are all kept arbitrary, and we have verified that all divergences are independent of them.

\section{Spin projections and Feynman diagrams}
\label{sec:method}

\subsection{Strategy}

Background gauge invariance implies that the one-loop divergence is a linear combination of the invariants in Eq.~\eqref{eq:action}. We use dimensional regularization in $d=4-2\epsilon$ with minimal subtraction. We can thus write
the divergent contributions to the effective action as
\begin{equation}
\Gamma^{(1)}_{\rm div}=-\frac{1}{2\epsilon}\!\int\!\dd^4x\sqrt g\Big[\beta_{1/\eta^2}C^2-\frac{\beta_{1/\xi^2}}{24}\hat R^2+\frac{\beta_{1/\alpha^2}}{4}\hat F^2\Big],
\label{eq:Gammadiv}
\end{equation}
with the coefficients being the beta functions of the inverse squared couplings, $\beta_{1/g^2}\equiv\dd(1/g^2)/\dd t=-\beta_{g^2}/g^4$ and $t\equiv\ln(\mu/\mu_0)$, where $\mu$ is the renormalization scale. Since the three invariants are independent, it is sufficient to compute the divergences on backgrounds where they can be told apart. We need only to consider two limits:
\begin{itemize}
\item[(a)] $\bar g_{\mu\nu}=\delta_{\mu\nu}+\bar h_{\mu\nu}$, $\bar\omega_\mu=0$. The two-point function of $\bar h_{\mu\nu}$ is sensitive to $C^2$ and to $\hat R^2\to R^2$;
\item[(b)] $\bar g_{\mu\nu}=\delta_{\mu\nu}$, $\bar\omega_\mu\neq0$. The two-point function of $\bar\omega_\mu$ is sensitive to $\hat F^2$ through its transverse part and to $\hat R^2\supset36(\partial_\mu\omega^\mu)^2$ through its longitudinal part.
\end{itemize}
The coefficient of $\hat R^2$ is therefore measured twice, which provides a nontrivial check of background Weyl invariance.\footnote{We stress that here $\bar h_{\mu\nu}$ is a perturbation of the background metric, not the quantum fluctuation $h_{\mu\nu}$.}

\subsection{Spin projectors}

In momentum space, the transverse and longitudinal projectors on vectors are~\cite{Percacci:2017fkn}
\begin{equation}
\label{eq:projectionSpin1}
P^T_{\mu\nu}=\delta_{\mu\nu}-\frac{p_\mu p_\nu}{p^2}\,,\qquad P^L_{\mu\nu}=\frac{p_\mu p_\nu}{p^2}\,,
\end{equation}
and on symmetric tensors
\begin{align}
P^{(2)}_{\mu\nu,\rho\sigma}&=\tfrac12\left(P^T_{\mu\rho}P^T_{\nu\sigma}+P^T_{\mu\sigma}P^T_{\nu\rho}\right)-\tfrac13P^T_{\mu\nu}P^T_{\rho\sigma}\,,\\
P^{(1)}_{\mu\nu,\rho\sigma}&=\tfrac12\left(P^T_{\mu\rho}P^L_{\nu\sigma}+P^T_{\mu\sigma}P^L_{\nu\rho}+P^T_{\nu\rho}P^L_{\mu\sigma}+P^T_{\nu\sigma}P^L_{\mu\rho}\right),\\
P^{(0,s)}_{\mu\nu,\rho\sigma}&=\tfrac13P^T_{\mu\nu}P^T_{\rho\sigma}\,,\qquad P^{(0,w)}_{\mu\nu,\rho\sigma}=P^L_{\mu\nu}P^L_{\rho\sigma}\,,\\
P^{(0,sw)}_{\mu\nu,\rho\sigma}&=\tfrac{1}{\sqrt3}P^T_{\mu\nu}P^L_{\rho\sigma}\,,\qquad P^{(0,ws)}_{\mu\nu,\rho\sigma}=\tfrac{1}{\sqrt3}P^L_{\mu\nu}P^T_{\rho\sigma}\,,
\end{align}
which satisfy orthogonality and completeness relations, $P^{(2)}+P^{(1)}+P^{(0,s)}+P^{(0,w)}=\mathbb I$, and have traces $5,3,1,1$. If $\Gamma^{(2)}(p)$ denotes the background two-point function, defined as the kernel of the quadratic part of the effective action, $\Gamma\supset\frac12\int\bar\Phi(-p)\,\Gamma^{(2)}(p)\,\bar\Phi(p)$, we define the normalized projections
\begin{align}
\Pi_2&=\tfrac15\,P^{(2)}\cdot\Gamma^{(2)}_{\bar h\bar h}\,,&
\Pi_0&=P^{(0,s)}\cdot\Gamma^{(2)}_{\bar h\bar h}\,,\nonumber\\
\Pi_T&=\tfrac13\,P^{T}\cdot\Gamma^{(2)}_{\bar\omega\bar\omega}\,,&
\Pi_L&=P^{L}\cdot\Gamma^{(2)}_{\bar\omega\bar\omega}\,,
\label{eq:projections}
\end{align}
where the dot denotes the full contraction of indices. Expanding the invariants to second order around flat space one finds (see also Appendix~\ref{app:projections}) that the divergence~\eqref{eq:Gammadiv} projects as
\begin{align}
\Pi_2^{\rm div}&=-\frac{p^4}{2\epsilon}\,\beta_{1/\eta^2}\,, &
\Pi_0^{\rm div}&=\frac{p^4}{8\epsilon}\,\beta_{1/\xi^2}\,,\nonumber\\
\Pi_T^{\rm div}&=-\frac{p^2}{2\epsilon}\,\beta_{1/\alpha^2}\,, &
\Pi_L^{\rm div}&=\frac{3p^2}{2\epsilon}\,\beta_{1/\xi^2}\,.
\label{eq:ansatz}
\end{align}
Each channel isolates one beta function. In particular, the Weyl-squared term contributes only to the spin-$2$ channel, the transverse channel is sensitive only to the kinetic term of the Weyl boson, and the $\hat R^2$ term controls both the spin-$0$ channel of the metric and the longitudinal channel of the Weyl vector. Background Weyl invariance implies the background Ward identity
\begin{equation}
\Pi_L^{\rm div}=\frac{12}{p^2}\,\Pi_0^{\rm div}\,,
\label{eq:ward}
\end{equation}
which follows from Eq.~\eqref{eq:Rhat} and is verified by our results.

\subsection{Tadpoles and bubbles}

\begin{figure}[t]
\centering
\figslot{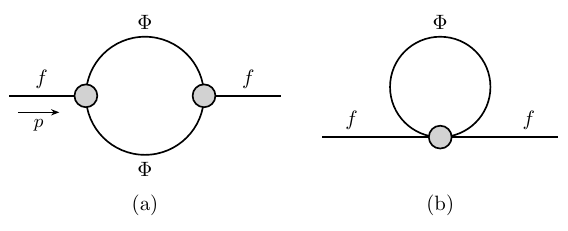}{One-loop topologies with two external background legs (wavy: $\bar h_{\mu\nu}$ or $\bar\omega_\mu$). Right: tadpole with a quartic vertex. Left: bubble with two cubic vertices. Internal lines: $h$, $v$, mixed $h$--$v$, ghosts $c,\bar c,\vartheta,\bar\vartheta$ (and mixed $c$--$\bar\vartheta$), third ghost $z$, matter.}
\caption{One-loop topologies contributing to the background two-point functions: bubbles (left) and tadpoles (right). The external legs $f$ are either two background gravitons $\bar h_{\mu\nu}$ (spin-$2$ and spin-$0$ channels) or two background Weyl vectors $\bar\omega_\mu$ (transverse and longitudinal channels). The internal lines $\Phi$ run over the graviton $h_{\mu\nu}$, the Weyl fluctuation $v_\mu$, the ghosts of diffeomorphisms  $\bar{c}_\mu,c^\mu$ and of Weyl  transformations $\bar{\vartheta},\vartheta$, the third ghost $z_\mu$ and the matter fields $\varphi$, including
the mixed propagators $\langle h v\rangle$ and $\langle c\bar\vartheta\rangle$.}
\label{fig:diagrams}
\end{figure}

The divergent parts of $\Pi_J$ are obtained from the one-loop diagrams with two external background legs, shown schematically in Fig.~\ref{fig:diagrams}. They are of two types: \emph{tadpoles}, built with one quartic vertex containing two background and two quantum fields, and \emph{bubbles}, built with two cubic vertices, each containing one background and two quantum fields. The vertices are obtained by expanding $S+S_{\rm gf}+S_{\rm gh}$ (and the matter action) to the required order in $\bar h_{\mu\nu}$ or $\bar\omega_\mu$ and in the fluctuations. The fields propagating in the loops are
\begin{itemize}
\item the graviton $h_{\mu\nu}$ and the Weyl fluctuation $v_\mu$, including the mixed propagator $\langle h\,v\rangle$ generated by the term $-12R\,\nabla_\mu\omega^\mu$ in $\hat R^2$;
\item the diffeomorphism ghosts $c^\mu,\bar c_\mu$, the Weyl ghosts $\vartheta,\bar\vartheta$, and the mixed ghost propagator $\langle c\,\bar\vartheta\rangle$;
\item the third ghost $z_\mu$, with kinetic operator $Y$;
\item the matter fields.
\end{itemize}

The mixed propagator $\langle h\,v\rangle$ is of order $\zeta$, where $\zeta$ is the gauge parameter introduced in the gauge fixing \eqref{eq:gfweyl}, but it combines with vertices of order $1/\zeta$ that arise from the background dependence of the Weyl gauge-fixing term itself, so that its contribution survives the Landau limit. In the diagrams containing these vertices, $\zeta$ is therefore kept finite and the limit $\zeta\to0$ is taken after the loop integration.

Denoting collectively by $\Phi$ the quantum fields, by $G=\langle\Phi\Phi\rangle$ their propagators and by $V^{(3)}$ and $V^{(4)}$ the vertices with one and two background legs, the two contributions read schematically
\begin{align}
\Gamma^{(2)}(p)&=\sum_{\rm tadpoles}s_T\int_q\mathrm{tr}\left[V^{(4)}(p,-p;q,-q)\,G(q)\right]\nonumber\\
&+\sum_{\rm bubbles}s_B\int_q\mathrm{tr}\left[V^{(3)}(p;q,-p-q)\,G(p+q)\right.\nonumber\\
&\qquad\qquad\left.\times V^{(3)}(-p;p+q,-q)\,G(q)\right],
\label{eq:diagrams}
\end{align}
where $\int_q=\int\dd^dq/(2\pi)^d$ and the trace runs over field species and indices. Each diagram is weighted by the appropriate symmetry factor $s_{T,B}$ ($1/2$ for loops of identical real bosons), by $-1$ for loops of Grassmannian fields, and with the opposite sign of a real bosonic loop for the third ghost, according to Eq.~\eqref{eq:Gamma1}. Since the number of contributions is large, we generated the list of diagrams and their weights automatically from the field content and the vertex structure. Individual diagrams depend on the gauge parameters; only their sum in each channel is gauge independent.

The loop integrals are dimensionally regularized. To separate ultraviolet from infrared divergences, all propagators are given an infrared regulator mass, $1/q^{2n}\to1/(q^2+m^2)^n$, which is removed at the end. After projection, the pole parts of the spin-$2$ and spin-$0$ channels are proportional to $p^4$, and those of the transverse and longitudinal channels to $p^2$, as required by Eq.~\eqref{eq:ansatz}.

\section{Beta functions}
\label{sec:betas}

\subsection{Results}

Collecting all diagrams, the divergences take the form~\eqref{eq:ansatz} with
\begin{align}
(4\pi)^2\,\beta_{1/\eta^2}&=\frac{810+\Nc}{120}\,,\label{eq:bres1}\\
(4\pi)^2\,\beta_{1/\alpha^2}&=\frac{2}{3}\,,\label{eq:bres2}\\
(4\pi)^2\,\beta_{1/\xi^2}&=-\frac{1}{48\,\xi^4}\Big[5\eta^4+1728\alpha^4\nonumber\\
&\quad+24(5\eta^2+24\alpha^2)\xi^2+16(10+\Nr)\xi^4\Big].\label{eq:bres3}
\end{align}
The one-loop beta functions of Weyl quadratic gravity coupled to matter are then
\begin{align}
(4\pi)^2\beta_{\eta^2} &= -\frac{810+N_C}{120}\,\eta^4 , \label{eq:betaeta}\\
(4\pi)^2\beta_{\alpha^2} &= -\frac{2}{3}\,\alpha^4 ,\label{eq:betaalpha} \\
(4\pi)^2\beta_{\xi^2} &= \frac{10+N_R}{3}\,\xi^4
  + \Big(\frac{5}{2}\,\eta^2 + 12\,\alpha^2\Big)\xi^2 \nonumber \\
  &\quad + \frac{5}{48}\,\eta^4 + 36\,\alpha^4 ,
  \label{eq:betaxi}
\end{align}
and represent the main new result of this work.
For pure gravity, $\Nc=\Nr=0$, so that $(4\pi)^2\beta_{\eta^2}=-\frac{27}{4}\eta^4$ and the first term of Eq.~\eqref{eq:betaxi} becomes $10\,\xi^4/3$. The matter content enters only through $\Nc$ and $\Nr$; in particular $\beta_{\alpha^2}$ is independent of ordinary matter.

The structure of Eqs.~\eqref{eq:betaeta}--\eqref{eq:betaxi} reflects the structure of the action. The beta functions of $\eta^2$ and $\alpha^2$ are decoupled from each other and from $\xi^2$. The beta function of $\xi^2$, instead, receives contributions from all couplings: the inhomogeneous terms $5\eta^4/48$ and $36\alpha^4$ show that the $\hat R^2$ coupling is generated by loops of the Weyl-squared and of the Weyl gauge sectors, the first being the same as in Riemannian quadratic gravity~\cite{Avramidi:1985ki,Codello:2006in}. The transverse channel is particularly simple: its pole,
\begin{equation}
\Pi_T^{\rm div}=-\frac{p^2}{3(4\pi)^2\,\epsilon}\,,
\end{equation}
does not depend on any coupling. For pure gravity the spin-$2$ pole is $\Pi_2^{\rm div}=-27p^4/[8(4\pi)^2\epsilon]$.

\subsection{\texorpdfstring{Asymptotic freedom of $\eta$ and $\alpha$}{Asymptotic freedom of eta and alpha}}

Equations~\eqref{eq:betaeta} and~\eqref{eq:betaalpha} integrate to
\begin{equation}
\begin{aligned}
\frac{1}{\eta^2(t)}&=\frac{1}{\eta_0^2}+\frac{810+\Nc}{120}\,\frac{t}{(4\pi)^2}\,,\\
\frac{1}{\alpha^2(t)}&=\frac{1}{\alpha_0^2}+\frac{2}{3}\,\frac{t}{(4\pi)^2}\,.
\end{aligned}
\label{eq:solutions}
\end{equation}
Both couplings are asymptotically free, with the logarithmic approach typical of marginal couplings.

For $\eta$ this is the well-known asymptotic freedom of the Weyl-squared coupling of quadratic gravity~\cite{Fradkin:1981iu,Avramidi:1985ki}. The coefficient $810+\Nc$ is the sum of the gravitational contribution, $798$ from the metric and ghosts as in the Riemannian theory plus $12$ from the Weyl vector, and of the matter contribution $\Nc$. For comparison, in conformal gravity, where the $R^2$ term is absent and Weyl transformations act on the metric alone, $798$ is replaced by $796$~\cite{Fradkin:1985am,OhtaPercacci:2015}; the origin of this difference is discussed in Ref.~\cite{Kawai:2025xbv}. Since every unitary field gives a positive contribution to $\Nc$ (it is proportional to the $c$ coefficient of the trace anomaly~\cite{Duff:1993wm,Vassilevich:2003xt}), matter makes $\eta$ run faster to zero. Asymptotic freedom of $\eta$ is lost only if $\Nc\le-810$, which requires a large number of fields with negative $\Nc$ such as the four-derivative scalars of Sec.~\ref{sec:HD}.

The asymptotic freedom of $\alpha$ is more surprising. For an Abelian gauge symmetry one would expect charged matter to screen the charge, as in quantum electrodynamics. In the present case the coefficient of the kinetic term of the Weyl boson grows linearly,
\begin{equation}
\frac{\dd}{\dd t}\left(\frac{1}{4\alpha^2}\right)=\frac{1}{6\,(4\pi)^2}=\frac{1}{96\pi^2}\,,
\label{eq:Kdot}
\end{equation}
with a coefficient that is independent of $\eta$, $\xi$, of the gauge, and of any matter with two-derivative kinetic terms. Two facts explain this behavior. First, matter does not contribute to the transverse channel at all: SM fermions and gauge fields do not couple to $\omega_\mu$, and the Higgs couples only to the longitudinal part of $\omega_\mu$, Eq.~\eqref{eq:higgsomega}. The same holds for any scalar with a two-derivative kinetic term, whose coupling to $\omega_\mu$ through the covariant derivative and through $\hat R$ is always, up to total derivatives, a coupling to the gradient $\partial_\mu\phi^2$. Second, the fields that carry Weyl charge and couple to the transverse Weyl boson belong to the gravitational sector: the graviton and the diffeomorphism ghosts carry charge 2, the Weyl boson interacts with them through the covariant derivatives and the curvatures, and these interactions are a direct consequence of the semidirect product structure of $\mathcal G$. In this sense the situation resembles a non-Abelian gauge theory, in which the gauge bosons are themselves charged and produce anti-screening. The net result of the gravitational loops is the anti-screening coefficient in Eq.~\eqref{eq:Kdot}, and the Weyl gauge coupling is asymptotically free for any matter content with two-derivative kinetic terms; four-derivative scalars can modify this conclusion, see Sec.~\ref{sec:HD}.

Combining Eqs.~\eqref{eq:solutions}, the ratio $\alpha^2/\eta^2$ approaches a universal value in the ultraviolet,
\begin{equation}
\frac{\alpha^2}{\eta^2}\xrightarrow{t\to\infty}\frac{3}{2}\cdot\frac{810+\Nc}{120}=\frac{810+\Nc}{80}\,,
\label{eq:ratioUV}
\end{equation}
for any initial condition with $\alpha_0\neq0$. For pure gravity this ratio is $81/8$. By Eq.~\eqref{eq:ratios_masses} this corresponds to $m_\omega^2/m_2^2\to3(810+\Nc)/10$.

\subsection{\texorpdfstring{The $\hat R^2$ coupling}{The R-hat squared coupling}}

The situation is different for $\xi$. For real couplings and $\Nr>-10$, all terms in Eq.~\eqref{eq:betaxi} are positive, and $\xi$ grows toward the UV. Keeping only the $\xi^4$ term, $(4\pi)^2\dd\xi^2/\dd t\ge(10+\Nr)\,\xi^4/3$, so that the one-loop solution diverges at a finite scale $t_L$ bounded by
\begin{equation}
t_L\le\frac{3(4\pi)^2}{(10+\Nr)\,\xi_0^2}=\frac{48\pi^2}{(10+\Nr)\,\xi_0^2}\,.
\label{eq:landaubound}
\end{equation}
The inhomogeneous terms in Eq.~\eqref{eq:betaxi} make the divergence occur earlier: even if $\xi_0\ll\eta_0$ at some scale, the ratio $\xi^2/\eta^2$ grows quickly, and $t_L$ is controlled by $\eta_0$ and $\alpha_0$. For instance, for $\eta_0=0.5$, $\alpha_0=0.1$ and any $\xi_0\lesssim10^{-2}$, numerical integration gives $t_L\simeq1.3\times10^3$. The hierarchy $\xi\ll\eta$ is therefore not stable under the RG flow toward the UV. The divergence itself lies outside the domain of validity of the one-loop approximation, since the loop expansion in the $\hat R^2$ sector is controlled by $\xi^2/(4\pi)^2$. However, by Eq.~\eqref{eq:landaubound} this parameter becomes of order one at most $\Delta t\simeq3/(10+\Nr)$ before $t_L$, so that $t_L$ reliably estimates the scale at which $\xi$ leaves the perturbative regime. Whether this signals a genuine Landau pole, or whether the flow can be continued through $1/\xi^2=0$~\cite{Salvio:2017qkx}, cannot be decided at one loop. The one-loop flow does determine, instead, whether asymptotically free trajectories exist in the physical region; to answer this question, in the next section we analyze its fixed-point structure.

\section{Renormalization group flow and fixed points}
\label{sec:flow}

\subsection{Coupling ratios and rescaled RG time}

The only common zero of Eqs.~\eqref{eq:betaeta}--\eqref{eq:betaxi} is the Gaussian fixed point, which is degenerate: since the beta functions are homogeneous of degree two in the squared couplings, the stability matrix vanishes there. The relevant question is therefore along which \emph{directions} the couplings can approach zero in the UV. Following the standard practice of quadratic gravity~\cite{Codello:2006in,Percacci:2017fkn}, we introduce the ratios
\begin{equation}
r\equiv\frac{\alpha^2}{\eta^2}\,,\qquad
w\equiv\frac{\eta^2}{16\,\xi^2}\,,
\label{eq:rw}
\end{equation}
and the rescaled RG time
\begin{equation}
\dd\tau=\frac{\eta^2}{(4\pi)^2}\,\dd t\,.
\label{eq:tau}
\end{equation}
The new variables resolve the degeneracy of the Gaussian fixed point: $\eta$ measures the distance from the origin, while $(r,w)$ parametrize the direction along which the origin is approached. The variable $w$ is regular at $\xi\to\infty$ ($w=0$) and coincides with the ratio $\omega$ of Ref.~\cite{Codello:2006in} when $\alpha=0$. By Eq.~\eqref{eq:ratios_masses}, $r=m_\omega^2/(24m_2^2)$ and $w=3m_2^2/(8\Lambda)$. Since $\eta^2>0$ decreases as $1/t$, $\tau$ grows monotonically with $t$, with $\tau\simeq[120/(810+\Nc)]\ln t$ for $t\to\infty$, so that the UV direction is preserved.

In these variables the flow becomes autonomous:
\begin{align}
\frac{\dd\eta^2}{\dd\tau}&=-\frac{810+\Nc}{120}\,\eta^2\,,\label{eq:etatau}\\
\frac{\dd r}{\dd\tau}&=r\left(\frac{810+\Nc}{120}-\frac23\,r\right),\label{eq:rtau}\\
\frac{\dd w}{\dd\tau}&=-\frac{1}{24}\left[A(r)\,w^2+B(r)\,w+\frac{10+\Nr}{2}\right],\label{eq:wtau}
\end{align}
with
\begin{equation}
A(r)=40+13824\,r^2\,,\qquad B(r)=222+\frac{\Nc}{5}+288\,r\,.
\label{eq:AB}
\end{equation}
The equations for $r$ and $w$ do not depend on $\eta$, so the flow in the $(r,w)$ plane is exact at one loop. The flow of $r$ does not depend on $w$: this is a consequence of the decoupling of $\beta_{\eta^2}$ and $\beta_{\alpha^2}$. Since both $w\to+\infty$ and $w\to-\infty$ correspond to $\xi\to0$, the variable $w$ lives on the projective line.

\subsection{Fixed points}

The fixed points of Eqs.~\eqref{eq:rtau}--\eqref{eq:wtau} are directions along which all couplings vanish with fixed ratios, i.e., asymptotically free trajectories. The equation $\dd r/\dd\tau=0$ has two solutions, which define two branches of fixed points:
\begin{equation}
\text{I:}\quad r^*=0\,,\qquad\text{II:}\quad r^*=\frac{810+\Nc}{80}\,.
\end{equation}
On branch I the Weyl boson is decoupled, $\alpha\equiv0$, while on branch II it has a nonvanishing gauge coupling. On each branch, $w^*$ solves a quadratic equation,
\begin{equation}
w^*_\pm=\frac{-B\pm\sqrt{\Delta}}{2A}\,,\qquad \Delta=B^2-2A(10+\Nr)\,,
\label{eq:wstar}
\end{equation}
with $A=40$, $B=222+\Nc/5$ on branch I, and $A=40+54(810+\Nc)^2/25$, $B=3138+19\Nc/5$ on branch II. There are therefore four fixed points. For pure gravity:
\begin{align}
\fp{{\rm I}\pm}:&\quad r^*=0\,,\quad w^*_\pm=\frac{-111\pm\sqrt{12121}}{40}\,,\\
\fp{{\rm II}\pm}:&\quad r^*=\frac{81}{8}\,,\quad w^*_\pm=\frac{-1569\pm i\sqrt{4624319}}{1417216}\,.
\end{align}
Branch I has two real fixed points, while on branch II the discriminant is negative and the two fixed points are complex conjugate. Their numerical values are collected in Table~\ref{tab:fp}.

\begin{table*}[t]
\caption{The four fixed points of the flow~\eqref{eq:rtau}--\eqref{eq:wtau} in pure gravity. The exponents $\kappa_r$ and $\kappa_w$ describe small deviations, $\delta r,\delta w\propto t^{\kappa}$ with $t=\ln(\mu/\mu_0)$; negative values are UV attractive. In all cases $\eta\propto t^{-1/2}$.}
\label{tab:fp}
\begin{ruledtabular}
\begin{tabular}{lccccl}
 & $r^*=\alpha^2/\eta^2$ & $w^*=\eta^2/16\xi^2$ & $\kappa_r$ & $\kappa_w$ & Properties \\
\hline
$\fp{{\rm I}+}$ & $0$ & $-0.02261$  & $+1$ & $-1.359$ & real, saddle: needs $\alpha\equiv0$ \\
$\fp{{\rm I}-}$ & $0$ & $-5.527$  & $+1$ & $+1.359$ & real, repulsive \\
$\fp{{\rm II}\pm}$ & $81/8$ & $-0.001107\pm0.001517\,i$  & $-1$ & --- & complex conjugate \\
\end{tabular}
\end{ruledtabular}
\end{table*}

\subsection{Stability}

The stability matrix of the system $(\eta^2,r,w)$ is triangular, since $\dd\eta^2/\dd\tau$ depends only on $\eta^2$ and $\dd r/\dd\tau$ does not depend on $w$, so that its eigenvalues are the diagonal entries. Since $\tau\simeq[120/(810+\Nc)]\ln t$, a perturbation that grows as $e^{\theta\tau}$ behaves as $t^{\kappa}$ in the RG time $t$, with $\kappa=120\,\theta/(810+\Nc)$. One finds
\begin{equation}
\begin{aligned}
\kappa_r&=\pm1 &&\text{branch I/II}\,,\\
\kappa_w&=\mp\frac{5\sqrt\Delta}{810+\Nc} &&\text{for }w^*_\pm\,,
\end{aligned}
\label{eq:exponents}
\end{equation}
while $\eta\propto t^{-1/2}$ along all fixed directions. Since $t=\ln(\mu/\mu_0)$, the deviations from the fixed directions scale as powers of $\ln\mu$, as expected for marginal couplings.

On branch I the $r$ direction is always UV repulsive: near $r=0$, $\alpha^2$ stays almost constant while $\eta^2$ decreases as $1/t$, so that $r$ grows linearly in $t$. Since $r=0$ is an invariant line, the fixed points of branch I are reached only by trajectories with $\alpha\equiv0$ exactly, i.e., with a completely decoupled Weyl boson. Along the line $r=0$ one has $\dd w/\dd\tau=-\frac{5}{3}(w-w^*_+)(w-w^*_-)$, so that $w$ decreases for $w>w^*_+$ and for $w<w^*_-$, and increases in between: on the projective line, $\fp{{\rm I}+}$ attracts all points except $\fp{{\rm I}-}$. Thus $\fp{{\rm I}+}$ is a saddle whose only attractive direction requires the decoupling of the Weyl interaction, while $\fp{{\rm I}-}$ is UV repulsive in both the $r$ and the $w$ directions; the radial direction $\eta$ is UV attractive at all fixed points. Both fixed points lie at $w^*<0$, i.e., $\xi^2<0$: they describe asymptotically free directions with a tachyonic $\hat R^2$ sector, in which the induced Planck mass would have the wrong sign. They are the counterpart of the two fixed points of Riemannian quadratic gravity, $\omega^*\simeq-0.0229$ and $\omega^*\simeq-5.467$~\cite{Avramidi:1985ki,Codello:2006in,Percacci:2017fkn}\fabrizio{\cite{OhtaPercacci:2013}}, which are recovered from our expressions by removing the Weyl vector, i.e., replacing $810\to798$; their tachyonic nature in the Riemannian case is well known~\cite{Salvio:2014soa,Salvio:2017qkx}.

On branch II the $r$ direction is UV attractive: every trajectory with $\alpha_0\neq0$ approaches $r=(810+\Nc)/80$, Eq.~\eqref{eq:ratioUV}. In pure gravity, however, $\Delta<0$ and the fixed points in $w$ are complex. On the line $r=(810+\Nc)/80$ the right-hand side of Eq.~\eqref{eq:wtau} then never vanishes and is always negative, so that $w$ decreases monotonically and reaches $w=0$, i.e., $\xi\to\infty$, at a finite RG time: this is the divergence of $\xi$ discussed in Sec.~\ref{sec:betas}. Branch II therefore carries no real fixed direction, and the trajectories with $\alpha_0\neq0$ are not asymptotically free in the $\hat R^2$ sector.

\subsection{The physical region}

In terms of the ratios, the physical region of Sec.~\ref{sec:physregion} is
\begin{equation}
r\ge0\,,\qquad w>0\,,
\end{equation}
with $w>1/16$ corresponding to $\xi<\eta$. None of the fixed points lies in it. More precisely, for $r\ge0$ and $w\ge0$ the coefficients $A(r)$ and $B(r)$ are positive (for $\Nc>-1110$), so that whenever $\Nr>-10$
\begin{equation}
\frac{\dd w}{\dd\tau}\le-\frac{10+\Nr}{48}<0\qquad\text{for }r\ge0,\ w\ge0\,.
\label{eq:nogo}
\end{equation}
Every trajectory starting in the physical region therefore reaches $w=0$ at a finite RG time, which is the divergence of $\xi$ discussed in Sec.~\ref{sec:betas}. No trajectory of the physical region reaches a fixed point: the vector sector is attracted to $r=(810+\Nc)/80$, but the $\hat R^2$ sector has no real zero there.

\begin{figure}[t]
\centering
\figslot{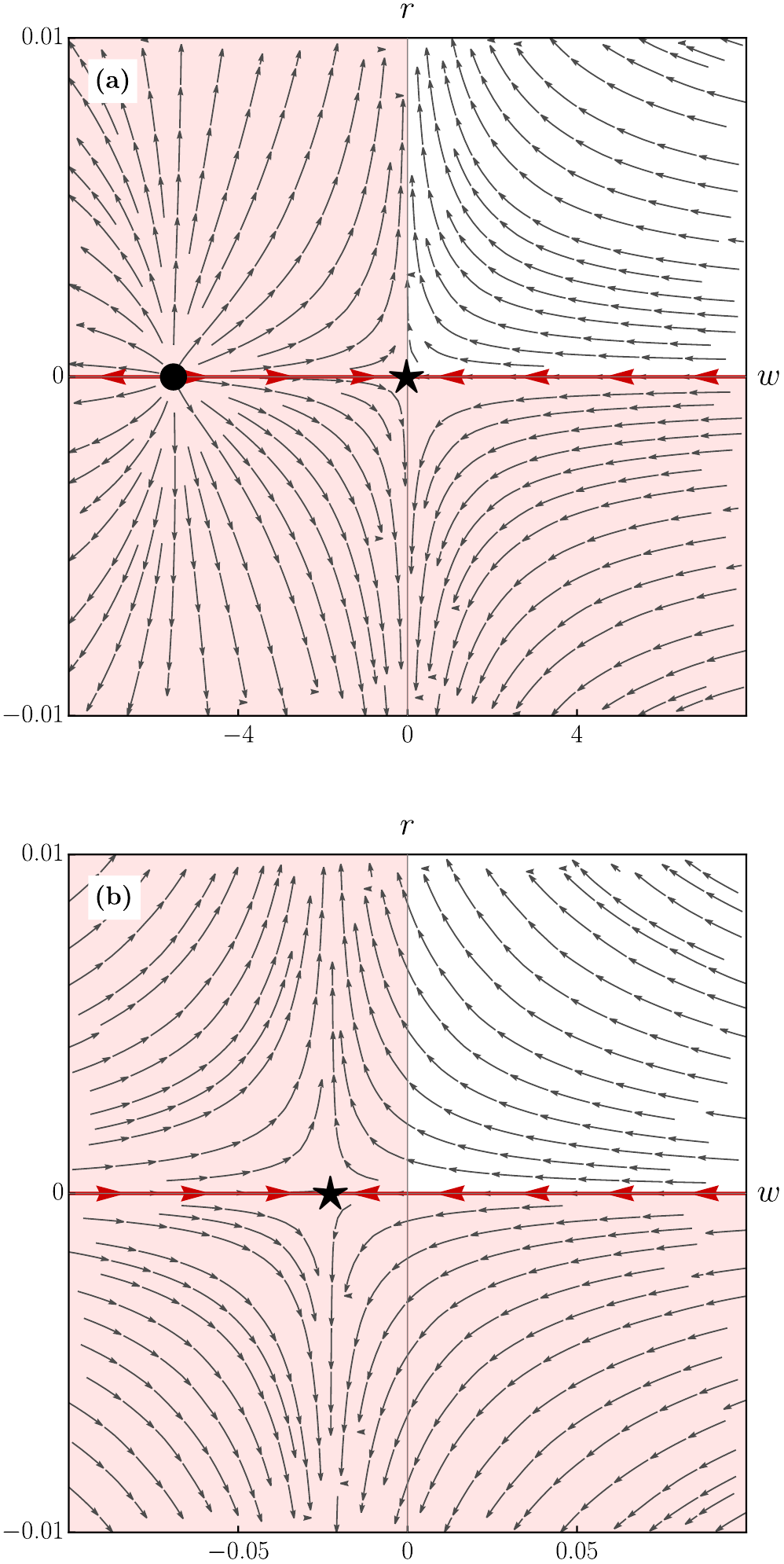}{Stream plot of Eqs.~(\ref{eq:rtau})--(\ref{eq:wtau}) in the plane $(w,r)$ for pure gravity. Mark $\mathrm{FP}_{{\rm I}\pm}$ on $r=0$ (green line, invariant), the line $r=81/8$ (dashed, UV attractive), the physical region $w>0$ (shaded) and the subregion $\xi<\eta$ ($w>1/16$). From notebook: \texttt{plotGlobalG}, \texttt{plotLocalG}.}
 \caption{RG flow of pure Weyl quadratic gravity in the plane
$(w,r)$, with $r=\alpha^2/\eta^2$ and $w=\eta^2/(16\xi^2)$; arrows point
toward the UV. (a) The real fixed points $\fp{{\rm I}-}$ (dot) and
$\fp{{\rm I}+}$ (star) lie on the line $r=0$ at $w<0$. (b) Zoom around
$\fp{{\rm I}+}$. Shaded regions are excluded (tachyonic); the white quadrant $r>0$, $w>0$ is the physical region.}
\label{fig:flowpure}
\end{figure}

\begin{figure}[t]
\centering
\figslot{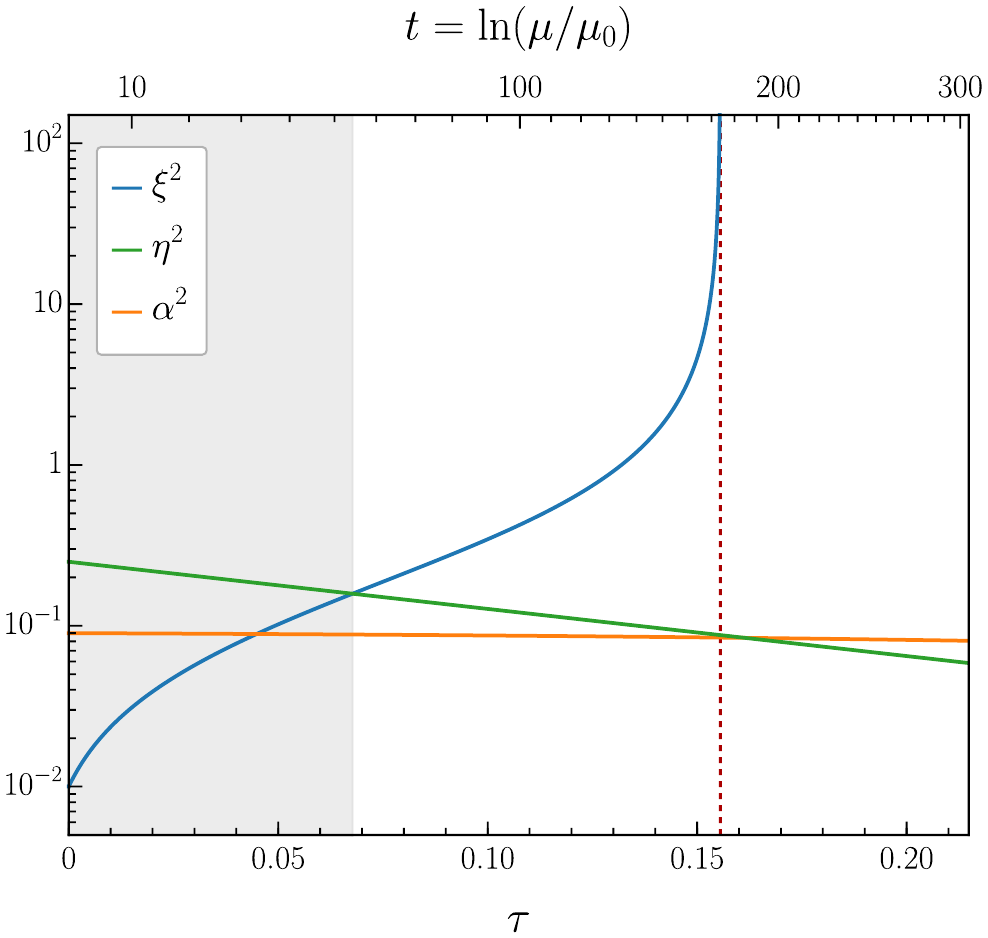}{Running of $\eta$, $\alpha$, $\xi$ (log scale) versus $\tau$ (bottom axis) and $t=\ln(\mu/\mu_0)$ (top axis) for an initial condition in the physical region, e.g.\ $(\eta_0,\alpha_0,\xi_0)=(0.5,0.3,0.1)$. Shade the intervals with $\xi<\eta$; mark the crossings $w=0$ ($\xi\to\infty$). From notebook: \texttt{plotRunningG}.}
\caption{
Running of $\eta^2$, $\xi^2$ and $\alpha^2$ in pure gravity along a trajectory starting in the physical region, $(\eta^2,\xi^2,\alpha^2)=(0.25,\,0.01,\,0.09)$, versus the rescaled RG time $\tau$ (bottom axis) and $t=\ln(\mu/\mu_0)$ (top axis). The couplings $\eta^2$ and $\alpha^2$ decrease, while $\xi^2$ grows and diverges at $w=0$ (dashed line), where it leaves the perturbative regime. The shaded band marks $\xi^2<\eta^2$, i.e., $w>1/16$.}
\label{fig:runningpure}
\end{figure}

\begin{figure}[t]
\centering
\figslot{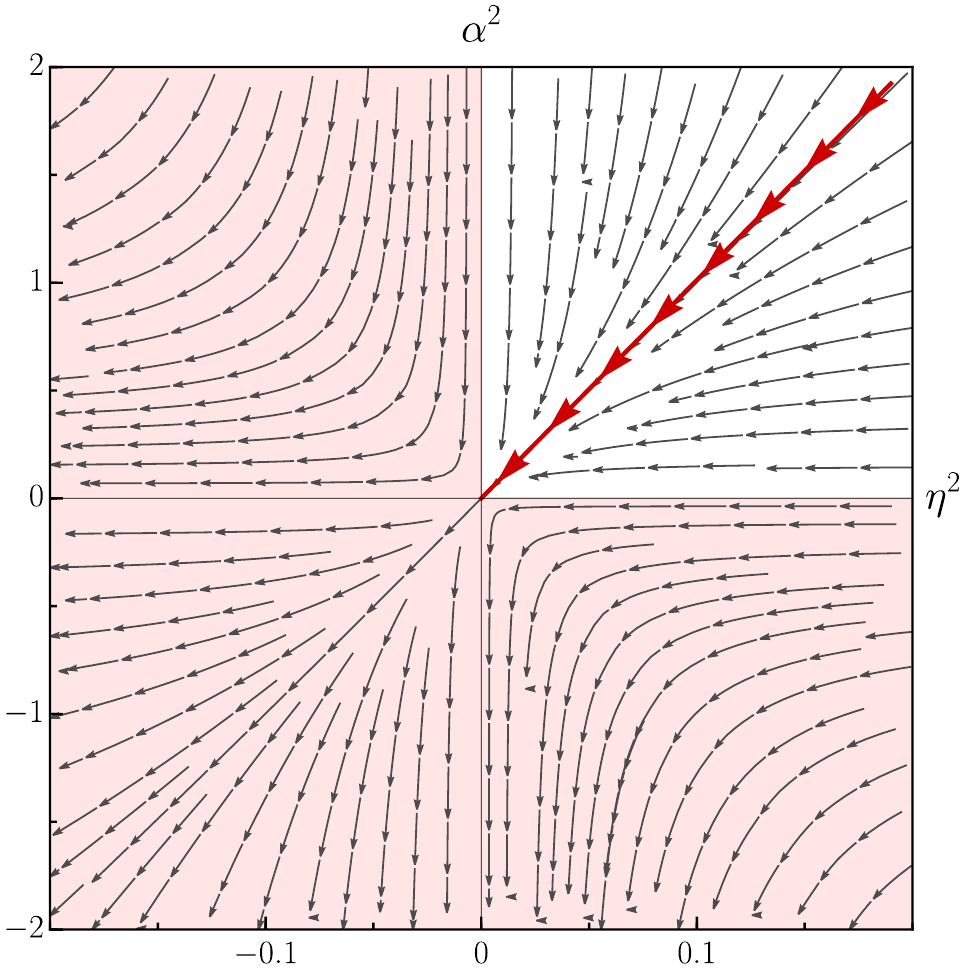}{Stream plot of the flow of Eqs.~(\ref{eq:betaeta})--(\ref{eq:betaalpha}) in the $(\eta,\alpha)$ plane: both couplings flow to zero; the UV-attractive direction $\alpha/\eta=\sqrt{81/8}=9/(2\sqrt2)$ and the repulsive direction $\alpha=0$. From notebook: section ``graphycs'', \texttt{myStreamPlot[\{betaEta,betaA\}]}.}
\caption{Flow of the Weyl-tensor coupling $\eta^2$ and of the Weyl gauge coupling $\alpha^2$, which is exact at one loop since $\beta_{\eta^2}$ and $\beta_{\alpha^2}$ do not depend on $\xi^2$. Both couplings are asymptotically free, and all trajectories with $\alpha^2\neq0$ approach the direction $\alpha^2/\eta^2=(810+\Nc)/80$ (red line, pure gravity: $81/8$). Shaded regions are excluded.}
\label{fig:gauge}
\end{figure}

The global structure of the flow is shown in Fig.~\ref{fig:flowpure}, and the behavior of the couplings along a typical physical trajectory in Fig.~\ref{fig:runningpure}. The flow in the $(\eta,\alpha)$ plane, which is exact since $\beta_{\eta^2}$ and $\beta_{\alpha^2}$ do not depend on $\xi$, is shown in Fig.~\ref{fig:gauge}.


\section{Standard Model matter}
\label{sec:SM}

\begin{figure}[t]
\centering
\figslot{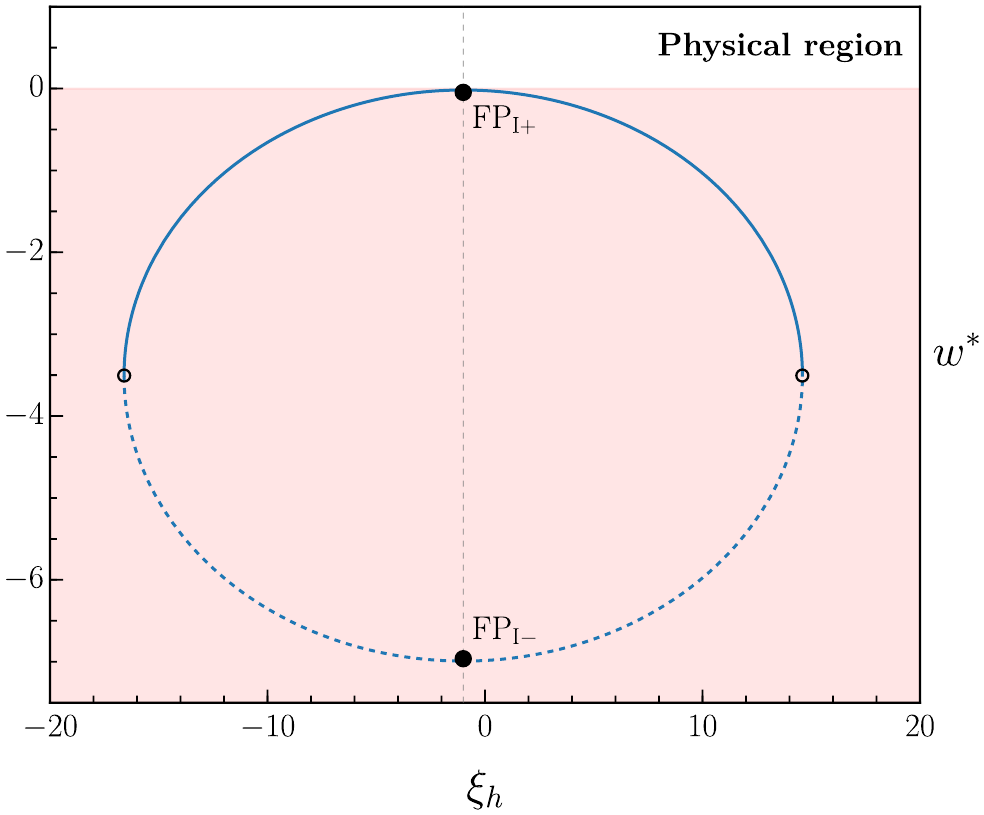}{SM, branch $r^*=0$: the two fixed points $w^*_\pm$ as functions of $\xi_h$ (conformal value $\xi_h=-1$). They stay below the excluded region $w<0$ (shaded) and collide at $|1+\xi_h|=15.59$.}
\caption{Fixed points of branch I ($\alpha=0$) for the SM as functions of the Higgs nonminimal coupling  $\xi_h$: $\fp{{\rm I}+}$ (solid) and
$\fp{{\rm I}-}$ (dashed). Both remain in the tachyonic region $w<0$ and collide (open circles) at $|1+\xi_h|=15.59$, beyond which they are complex. The vertical line marks the conformal value $\xi_h=-1$. The excluded region $w<0$ is shaded.}
\label{fig:SM}
\end{figure}

The effect of matter on the fixed points of Riemannian quadratic gravity has been studied in Refs.~\cite{Salvio:2014soa,Salvio:2017qkx,Hamada:2017rvn}. In our case the matter content enters the flow only through $\Nc$ and $\Nr$, Eq.~\eqref{eq:NcNr}; for the SM, $\Nc=292$ and $\Nr=4(1+\xi_h)^2$. Matter has the following effects.
\begin{itemize}
\item $\beta_{\alpha^2}$ is unchanged, and $\eta$ runs faster to zero. The attractive line of branch II moves to $r^*=(810+\Nc)/80=13.775$, i.e., $m_\omega^2/m_2^2\to3(810+\Nc)/10\simeq331$.
\item On branch I the two real fixed points move with $\xi_h$, but remain at $w^*<0$ for any $\xi_h$, in agreement with Eq.~\eqref{eq:nogo}. Matter with positive $\Nc$ moves $\fp{{\rm I}+}$ toward $w=0$, while a deviation of $\xi_h$ from the conformal value $-1$ pushes it away. The two fixed points collide and become complex for $|1+\xi_h|>15.59$; beyond this value branch I has no real fixed points either.
\item On branch II the fixed points remain complex for any $\xi_h$.
\end{itemize}
Representative values are given in Table~\ref{tab:matter}, and the dependence of the branch-I fixed points on $\xi_h$ is shown in Fig.~\ref{fig:SM}.

More generally, the no-go statement~\eqref{eq:nogo} relies only on $\Nc>-1110$ and $\Nr>-10$. Asymptotic freedom of $\eta$ requires $\Nc>-810$, and every field with a standard two-derivative kinetic term gives a non-negative contribution to $\Nr$: $(1+\xi_i)^2$ for scalars and zero for fermions and gauge vectors. Hence we can make a general statement:

\emph{For any matter content made of scalars, fermions and gauge vectors with two-derivative kinetic terms, including the SM for any value of $\xi_h$, Weyl quadratic gravity has no ultraviolet fixed point in the physical region at one loop: the asymptotically free fixed directions either require $\alpha\equiv0$ and $\xi^2<0$ (branch I), or are complex (branch II), and along every physical trajectory $\xi$ leaves the perturbative regime at a finite scale.}

The obstruction is thus entirely in the $\hat R^2$ sector, and Eq.~\eqref{eq:wtau} shows what is needed to remove it: a negative constant term, i.e., $\Nr<-10$. In that case the product of the roots $(10+\Nr)/(2A)$ is negative, and on both branches one fixed point has $w^*>0$.

\begin{table*}[t]
\caption{Fixed points for different matter contents. $\Nc$ and $\Nr$ are defined in Eqs.~\eqref{eq:NcNr} and~\eqref{eq:NcNrHD}; ``P'' denotes four-derivative scalars at the Paneitz point and $\xi_h$ the Higgs nonminimal coupling ($\xi_h=-1$ unless indicated), while $\hat{\square}^2$ denotes the minimal four-derivative scalar of Eq.~\eqref{eq:SHD} with
$\xi_{1,2}=\lambda_i=0$ for $i\neq3$, for which $r^*_{\rm II}$ is given by
Eq.~\eqref{eqall:rstar_HD}. Branch I has $r^*=0$. A fixed point is physical if $r^*\ge0$ and $w^*>0$.}
\vspace{4pt}
\label{tab:matter}
\begin{ruledtabular}
\begin{tabular}{lrrcccl}
Matter & $\Nc$ & $\Nr$ & $w^*_{{\rm I}\pm}$ & $r^*_{\rm II}$ & $w^*_{{\rm II}\pm}$ & Physical FP \\
\hline
Pure gravity & 0 & 0 & $-0.0226$, $-5.527$ & 10.125 & $(-1.11\pm1.52\,i)\times10^{-3}$ & none \\
SM & 292 & 0 & $-0.0179$, $-6.992$ & 13.775 & $(-0.81\pm1.12\,i)\times10^{-3}$ & none \\
SM, $\xi_h=0$ & 292 & 4 & $-0.0251$, $-6.985$ & 13.775 & $(-0.81\pm1.42\,i)\times10^{-3}$ & none \\
SM $+$ 36 P, $\lambda_3=0$ & 4 & 0 & $-0.0225$, $-5.547$ & 10.175 & $(-1.10\pm1.51\,i)\times10^{-3}$ & none \\
SM $+$ 136 P, $\lambda_3=0$ & $-796$ & 0 & $-0.0841$, $-1.486$ & 0.175 & $-0.0579$, $-0.186$ & none \\
SM $+$ 36 P, $\lambda_3=0.01$ & 4 & $-25.92$ & $+0.0355$, $-5.606$ & 10.175 & $+1.50\times10^{-3}$, $-3.70\times10^{-3}$ & $\fp{{\rm II}+}$ \\
SM + 36\,$\hat\Box^2$, $\lambda_3 = 0$ & 364 & 72 & $-0.1418$, $-7.228$ & 0.397
  & $(-0.92 \pm 1.00\,i)\times 10^{-1}$ & none \\
SM + 36\,$\hat\Box^2$, $\lambda_3 = 0.04$ & 364 & $-31.68$ & $+0.0366$, $-7.407$ & 0.397
  & $+2.35\times 10^{-2}$, $-0.208$ & FP$_{\rm II+}$ \\
\end{tabular}
\end{ruledtabular}
\end{table*}

\section{Higher-derivative scalars}
\label{sec:HD}

\subsection{Weyl-invariant dimensionless scalars}

Fields with $\Nr<0$ cannot have a standard two-derivative kinetic term. A natural candidate in a scale-invariant setting is a dimensionless scalar $\varphi$ with a four-derivative kinetic term~\cite{Fradkin:1985am,Paneitz:2008afy,Boyle:2021jaz,Buccio:2024omv}. In Weyl geometry, a scalar of zero Weyl charge with action
\begin{equation}
\begin{split}
S_{\rm HD}&=\frac12\int\dd^4x\sqrt g\;\varphi\,\hat{\mathcal D}_4\,\varphi\,,\\
\hat{\mathcal D}_4&=\hat\nabla^2\hat\nabla^2
+\hat\nabla_\mu\left(\xi_1\hat R^{(\mu\nu)}+\xi_2\,g^{\mu\nu}\hat R\right)\hat\nabla_\nu\\
&\quad+\lambda_1C^2+\lambda_2\hat R_{(\mu\nu)}\hat R^{(\mu\nu)}+\lambda_3\hat R^2\\
&\quad+\lambda_4\hat\square\hat R+\lambda_5\hat F^2
\end{split}
\label{eq:SHD}
\end{equation}
is Weyl invariant for any value of the couplings, because all ingredients transform covariantly. For $\xi_i=0$, $\lambda_i=0$, we obtain a minimal higher-derivative scalar with kinetic term $\hat{\square}^2$.
At the Paneitz point~\cite{Paneitz:2008afy} $\xi_1=2$, $\xi_2=-2/3$, $\lambda_i=0$, the operator reduces, up to total derivatives, to the Riemannian Paneitz operator, which is Weyl covariant on its own: the scalar then decouples from $\omega_\mu$. This is a special case of the criterion of Ref.~\cite{Iorio:1996ad}, according to which a theory is conformally invariant if it admits a
Weyl gauging in which $\omega_\mu$ enters only through
$\hat F_{\mu\nu}$, up to total derivatives; here this dependence vanishes identically.

Away from the Paneitz point, the coupling of $\varphi$ with $\omega_\mu$ is realized through the deviations of the couplings to the Weyl-covariant curvatures from their Paneitz values. In what follows we deform the Paneitz point only by the coupling $\lambda_3$ to $\hat R^2$, which, like a nonconformal coupling of the Higgs to $\hat R$, couples $\omega_\mu$ through $\nabla_\mu\omega^\mu$ and $\omega^2$.
This is the minimal displacement from the conformal point that makes the asymptotically free fixed point appear. We will briefly discuss how  $\lambda_5\ne0$, as well as the minimal scalar case with $\xi_{i}=0$, can reshape the space where this fixed point becomes real, but not move the bound for it to be in the physical region.

Each such scalar contributes to the one-loop divergences with weights
\begin{align}
w_C&=\tfrac52(\xi_1-2)^2-8-120\lambda_1-60\lambda_2\,,\label{eq:wC}\\
w_R&=\tfrac52(\xi_1-2)^2+18(\xi_1-2)\big(\xi_2+\tfrac23\big)+36\big(\xi_2+\tfrac23\big)^2\nonumber\\
&\quad-24\lambda_2-72\lambda_3\,,\label{eq:wR}\\
w_T&=\tfrac16(\xi_1-2)^2-4\lambda_2-4\lambda_5\,,\label{eq:wT}
\end{align}
where $w_C$ and $w_R$ are in units of a minimally coupled scalar, while
$w_T$ is the contribution to $(4\pi)^2\beta_{1/\alpha^2}$, to be compared
with the gravitational value $2/3$ in Eq.~\eqref{eq:bres2}. The weights
$w_C$ and $w_R$ shift those of Eq.~\eqref{eq:NcNr},
\begin{equation}
N_C\to N_C+N_{\rm HD}\,w_C\,,\qquad
N_R\to N_R+N_{\rm HD}\,w_R\,.
\label{eq:NcNrHD}
\end{equation}
In what follows, $N_C$ and $N_R$ denote these total weights, so that the beta functions become
\begin{align}
(4\pi)^2\beta_{\eta^2} &= -\frac{810+N_C}{120}\,\eta^4\,,
\label{eq:betaeta_HD}\\
(4\pi)^2\beta_{\alpha^2} &= -\Big(\frac{2}{3}+N_{\rm HD}\,w_T\Big)\alpha^4\,,
\label{eq:betaalpha_HD}\\
(4\pi)^2\beta_{\xi^2} &= \frac{10+N_R}{3}\,\xi^4
+\Big(\frac{5}{2}\,\eta^2+12\,\alpha^2\Big)\xi^2\nonumber\\
&\quad+\frac{5}{48}\,\eta^4+36\,\alpha^4\,.
\label{eq:betaxi_HD}
\end{align}
The flow of the ratios, Eqs.~\eqref{eq:rtau}--\eqref{eq:wtau}, becomes
\begin{align}
\frac{\mathrm dr}{\mathrm d\tau} &= r\left(\frac{810+N_C}{120}
-\Big(\frac{2}{3}+N_{\rm HD}\,w_T\Big)r\right),
\label{eq:rflow_HD}\\
\frac{\mathrm dw}{\mathrm d\tau} &= -\frac{1}{24}\left[A(r)\,w^2+B(r)\,w
+\frac{10+N_R}{2}\right],
\label{eq:wflow_HD}
\end{align}
with $A(r)$ and $B(r)$ as in Eq.~\eqref{eq:AB}. 

The weights $w_C$ and $w_R$ agree with Ref.~\cite{Buccio:2024omv}.
At the Paneitz point $w_C=-8$, the value of Fradkin and Tseytlin~\cite{Fradkin:1985am}, while $w_R=w_T=0$. The quadratic form in $(\xi_1-2,\xi_2+2/3)$ in Eq.~\eqref{eq:wR} is positive definite: without the couplings $\lambda_2$, $\lambda_3$, higher-derivative scalars cannot make $\Nr$ negative. The coupling $\lambda_3$ does not enter the transverse channel, since $\varphi^2\hat R^2$ couples $\omega_\mu$ only through $\nabla_\mu\omega^\mu$ and $\omega^2$; with $\lambda_5=\lambda_2=0$ at the Paneitz values of $\xi_1,\xi_2$ the transverse channel receives no contribution and $\beta_{\alpha^2}$ is unchanged. The weights do not depend on $\lambda_4$: for vanishing scalar background, $\hat\square\hat R$ is linear in the curvature and, inserted in a tadpole, gives a total derivative, so that it does not
contribute to the $\mu$-running, although it does contribute to the physical running through bubble diagrams~\cite{Buccio:2024omv}. The spin-$0$ contributions of Eq.~\eqref{eq:wR} are computed on backgrounds with $\bar\omega_\mu=0$, where $\hat R=R$; the corresponding longitudinal contributions follow from the Ward identity~\eqref{eq:ward}.

\subsection{Paneitz scalars}

With $\NHD$ scalars at the Paneitz point, $\Nc=\Nc^{\rm SM}-8\NHD$ and $\Nr$ is unchanged. Paneitz scalars slow down the running of $\eta$, which loses asymptotic freedom for $\NHD\ge138$ when added to the SM. The line of branch II moves toward the origin, $r^*=(810+\Nc)/80$, since $w_T=0$ at the Paneitz point, and its fixed points become real only very close to the loss of asymptotic freedom, for $\NHD=136$ and $137$ with a conformally coupled Higgs. In this case the fixed point $\fp{{\rm II}+}$ is UV attractive in all directions, but it still lies at $w^*<0$ (see Table~\ref{tab:matter}). In agreement with the no-go result~\eqref{eq:nogo}, Paneitz scalars can make the fixed points real, but cannot bring them into the physical region.

\subsection{A physical, fully attractive fixed point}

The situation changes with a positive coupling $\lambda_3$. For $\NHD$ scalars at the Paneitz point with coupling $\lambda_3$, added to the SM,
\begin{equation}
\label{eq:weightsHD}
\Nc=292-8\NHD\,,\qquad \Nr=4(1+\xi_h)^2-72\,\NHD\,\lambda_3\,.
\end{equation}
As $\lambda_3$ grows, $\Nr$ decreases, and the following happens on branch II:
\begin{enumerate}
\item for $10+\Nr\le B^2/(2A)$ the complex pair $\fp{{\rm II}\pm}$ collides on the real axis and becomes a pair of real fixed points, both still at $w^*<0$;
\item for $\Nr<-10$ the fixed point $w^*_+$ crosses $w=0$, i.e., passes through $\xi^*\to\infty$, and enters the physical region.
\end{enumerate}
In addition, $r^*>0$ is required. As the next equation shows, for scalars at the Paneitz point this condition coincides with the asymptotic freedom of $\eta^2$.
The resulting fixed point $\fp{{\rm II}+}$ has
\begin{equation}
\begin{aligned}
r^*&=\frac{810+\Nc}{80}>0\,,\\
w^*_+&=\frac{\sqrt{B^2+2A|10+\Nr|}-B}{2A}>0\,,
\end{aligned}
\end{equation}
and exponents $\kappa_r=-1$, $\kappa_w=-5\sqrt\Delta/(810+\Nc)$, together with $\eta\propto t^{-1/2}$: it is UV attractive in all directions. Moreover, it has a nonvanishing Weyl gauge coupling, $\alpha^{*2}/\eta^{*2}=(810+\Nc)/80$, i.e., the Weyl boson does not decouple. Since for $\Nr<-10$ one has $\dd w/\dd\tau=-(10+\Nr)/48>0$ at $w=0$, the physical region $\{r>0,w>0\}$ is invariant under the flow toward the UV, and it is entirely contained in the basin of attraction of $\fp{{\rm II}+}$. On branch I, $w^*_+$ also becomes positive, but this fixed point remains a saddle reached only for $\alpha\equiv0$.
As a benchmark we take $\NHD=36$, the number of dimension-zero scalars required to cancel the Weyl anomaly of the SM in the proposal of Ref.~\cite{Boyle:2021jaz}, and a conformally coupled Higgs, $\xi_h=-1$. Then $\Nc=4$ and
\begin{equation}
\begin{aligned}
\lambda_3&>2.52\times10^{-3}:&&\fp{{\rm II}\pm}\ \text{real}\,,\\
\lambda_3&>3.86\times10^{-3}:&&\fp{{\rm II}+}\ \text{physical}\,.
\label{eq:lambench}
\end{aligned}
\end{equation}
For example, $\lambda_3=0.01$ gives $r^*=10.175$, $w^*_+=1.50\times10^{-3}$, i.e., $\xi^*/\eta^*\simeq6.5$, and $\kappa_w\simeq-46$. The general condition is
\begin{equation}
72\,\NHD\,\lambda_3>10+4(1+\xi_h)^2\,.
\label{eq:lambd}
\end{equation}

\begin{figure}[t!]
\centering
\figslot{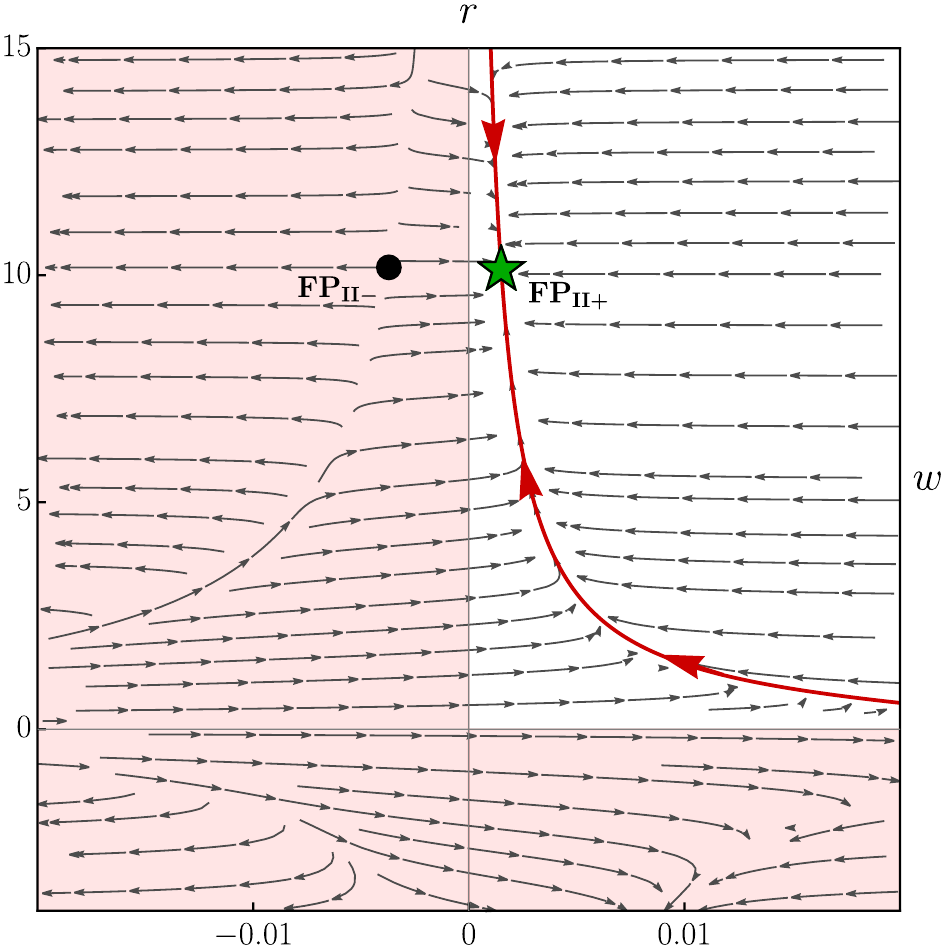}{Same as Fig.~\ref{fig:flowpure} for SM $+$ 36 Paneitz scalars with $\lambda_3=0.01$, $\xi_h=-1$: $\mathrm{FP}_{{\rm II}+}$ at $(r^*,w^*)=(10.175,\,1.5\times10^{-3})$ inside the physical region, attracting all trajectories with $r>0$, $w>0$.}
\caption{RG flow in the $(w,r)$ plane for the SM with a conformally coupled Higgs ($\xi_h=-1$) and 36 Weyl-invariant four-derivative scalars with $\lambda_3=0.01$. The fixed point $\fp{{\rm II}+}$ lies in the physical region and attracts all trajectories with $\alpha\neq0$ and real $\xi$.}
\label{fig:flowHD}
\end{figure}

\begin{figure}[t!]
\centering
\figslot[\columnwidth]{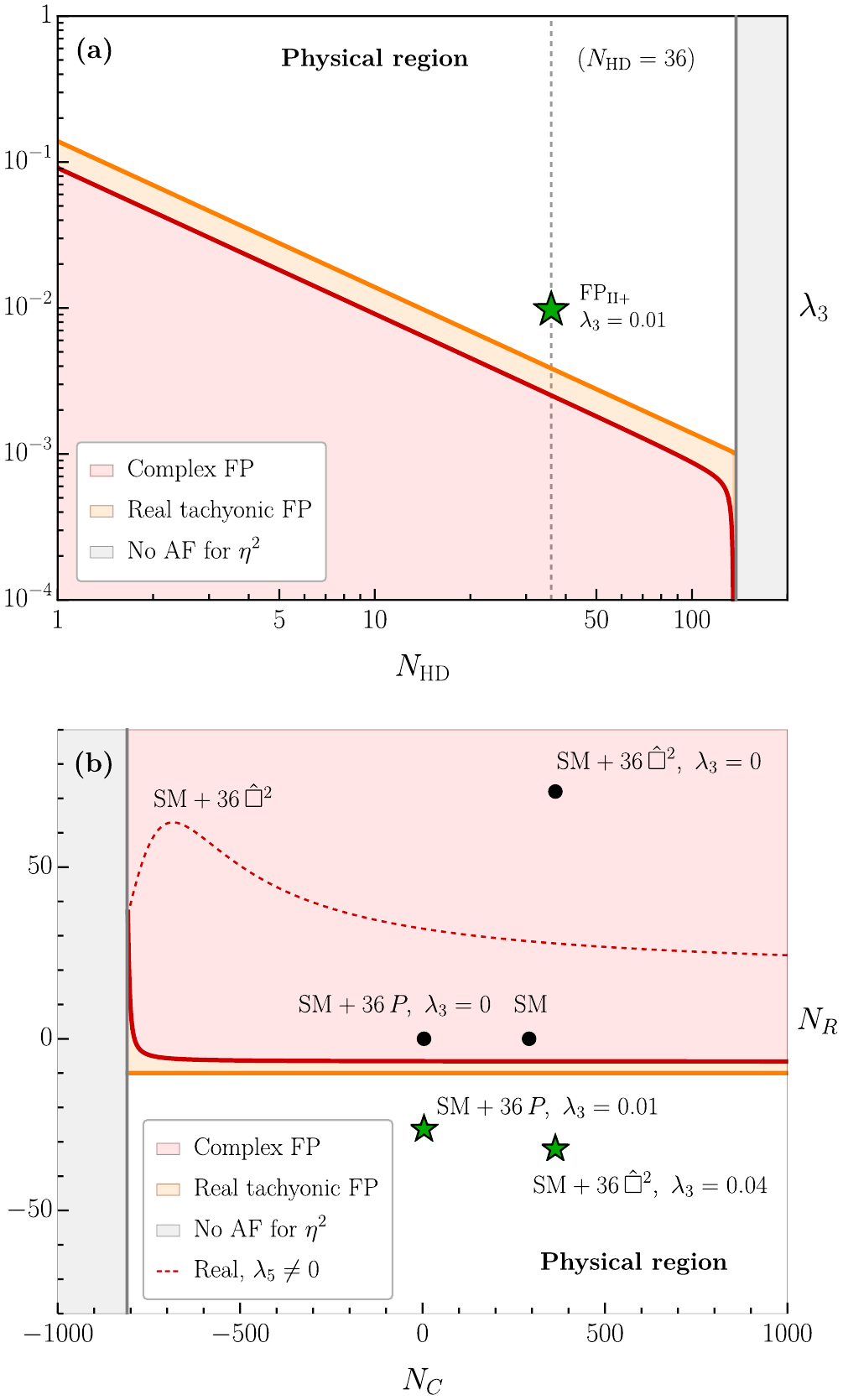}{Regions in the $(N_{\rm HD},\lambda_3)$ plane ($\xi_h=-1$): (i) $\mathrm{FP}_{{\rm II}\pm}$ complex; (ii) real but tachyonic ($0<10+N_R\le B^2/2A$); (iii) $\mathrm{FP}_{{\rm II}+}$ physical ($N_R<-10$, curve $72N_{\rm HD}\lambda_3=10$); (iv) no asymptotic freedom ($N_{\rm HD}\ge138$).}
\caption{Fixed-point structure of the $\fp{{\rm II}+}$ solution for higher-derivative scalars coupled with $\lambda_3\neq0$.
Excluded regions are of two kinds: tachyonic with a real fixed-point solution (orange) and complex solution (red). The physical region is white, and the gray area is where the AF of $\eta^2$ is lost.
(a) Plane $(\NHD,\lambda_3)$ of the number $\NHD$ of four-derivative scalars and of their coupling $\lambda_3$ to $\hat R^2$, for the SM with $\xi_h=-1$. The dotted line shows the benchmark case $\NHD=36$, and the star the fixed point for $\lambda_3=0.01$.
(b) Plane $(N_C,N_R)$ of the variables defined in Eq.~\eqref{eq:NcNrHD}. The dots indicate the positions of complex solutions for the theories indicated, while the stars indicate physical fixed points. The solid lines represent the bounds for $w_T=0$;  the dashed red line shows how the reality bound (red line) changes for 36 $\hat\square^2$ scalars ($\lambda_5=0$, $w_T=2/3$); this curve applies to the $\hat\square^2$ points.
The transverse weight $w_T$ moves the reality bound, while leaving the AF bound (gray line) and the physical one (orange line) untouched.}
\label{fig:phaseHD}
\end{figure}

The fixed point lies at $\xi^*>\eta^*$; reaching $w^*>1/16$, i.e., $\xi^*<\eta^*$, would require $\lambda_3=\order(1)$. This is not in contradiction with the phenomenological requirement $\xi\ll\eta$, which concerns the infrared values of the couplings along the trajectory and not the UV ratios. The flow is shown in Fig.~\ref{fig:flowHD} and the regions in the $(\NHD,\lambda_3)$ plane in Fig.~\ref{fig:phaseHD}.

When $w_T\neq0$, as for Paneitz scalars with $\lambda_5\neq0$ or for $\hat\square^2$ scalars, two new features appear:
\begin{enumerate}
    \item the transverse weight in the flow of $r$ moves the attractive line
of branch~II to
\begin{equation}
\tilde{r}^*=\frac{810+N_C}{80+120\,N_{\rm HD}\,w_T}\,.
\label{eqall:rstar_HD}
\end{equation}
As a consequence, in this case the
positivity of $\tilde{r}^*$ no longer coincides with the AF of $\eta^2$, which remains $N_\mathrm{C}>-810$. 
For $\hat\square^2$ scalars, $w_T=2/3>0$: the asymptotic freedom of $\alpha$ is reinforced and $\tilde r^*<(810+N_C)/80$. For scalars at the Paneitz point deformed by $\lambda_5$, $w_T=-4\lambda_5$: then $\tilde r^*>(810+N_C)/80$ as long as $\alpha$ remains asymptotically free, i.e., $\lambda_5<1/(6N_{\rm HD})$; beyond this value $\tilde r^*<0$ and branch~II disappears.
\item the reality condition $\Delta\ge0$, Eq.~\eqref{eq:wstar}, i.e.,
$10+N_R\le B^2/(2A)$, is evaluated with $A(r)$ and $B(r)$ of
Eq.~\eqref{eq:AB} at $r=\tilde r^*$: $w_T$ moves the reality bound,
while the physical bound $10+N_R<0$ is unchanged (see Fig.~\ref{fig:phaseHD}).
\end{enumerate}
For 36 minimal $\hat\Box^2$ scalars and the SM matter with $\xi_h = -1$,
$\tilde r^* \simeq 0.397$, FP$_{\rm II\pm}$ become real for
$\lambda_3 > 1.71\times 10^{-2}$, while FP$_{\rm II+}$ is physical for
$\lambda_3 > 3.16\times 10^{-2}$, to be compared with
$2.52\times 10^{-3}$ and $3.86\times 10^{-3}$ for Paneitz scalars,
Eq.~\eqref{eq:lambench}. The resulting fixed point is again UV
attractive in all directions, but more weakly than for Paneitz
scalars, because of the smaller value of $\tilde r^*$ (see Table~\ref{tab:matter}).

\subsection{Caveats}

This mechanism should be regarded as a proof of principle. First, we have treated the couplings $\xi_1$, $\xi_2$, $\lambda_i$ of the scalars as fixed parameters; a complete analysis requires their beta functions, and in particular the verification that the Paneitz point deformed by $\lambda_3$ is consistent with the flow. Second, four-derivative scalars contain negative-norm states, as does the spin-$2$ sector of quadratic gravity; their presence does not worsen the unitarity problem of the theory, but does not solve it either. Third, our beta functions describe the dependence on the subtraction scale $\mu$. In the physical, momentum-dependent running of Refs.~\cite{Buccio:2024hys,Buccio:2024omv}, the contributions of $\lambda_1$, $\lambda_2$ and $\lambda_3$ to the gravitational beta functions are absent, and the mechanism would need to be reconsidered.

\section{Discussion and conclusions}
\label{sec:conclusions}

We have computed the one-loop beta functions of Weyl quadratic gravity, the gauge theory of diffeomorphisms and dilatations formulated in Weyl conformal geometry, including the SM and Weyl-invariant four-derivative scalars. The calculation was performed diagrammatically in the background field method, with a gauge-fixing adapted to the semidirect product structure of the gauge group, the full ghost sector including the third ghost, and a projection of the background two-point functions onto spin components. The results pass a number of nontrivial checks, among which the reduction to Riemannian quadratic gravity plus a Maxwell field when the Weyl gauge coupling vanishes, the Ward identity between the spin-$0$ graviton and longitudinal Weyl channels, and gauge independence.

Our main findings are the following.
\begin{enumerate}
\item The Weyl-tensor coupling $\eta$ and the Weyl gauge coupling $\alpha$ are asymptotically free. The running of $\alpha$, Eq.~\eqref{eq:Kdot}, is universal: it does not depend on the other couplings nor on ordinary matter, and it is due to the anti-screening effect of the gravitational sector, which is charged under dilatations. Asymptotically, $\alpha^2/\eta^2\to(810+\Nc)/80$, so that in the broken phase $m_\omega^2/m_2^2\to3(810+\Nc)/10$.
\item The $\hat R^2$ sector is the obstruction to an ultraviolet completion. The flow of the coupling ratios has four fixed points: two real ones, which require $\alpha\equiv0$ and $\xi^2<0$ and are, respectively, a saddle and a fully repulsive point, and two complex conjugate ones. In the physical region $\xi$ always leaves the perturbative regime at a finite scale.
\item This conclusion holds for any matter content with non-negative weight in the $\hat R^2$ divergence, including the SM with any Higgs nonminimal coupling.
\item Weyl-invariant dimensionless scalars with a positive coupling $\lambda_3$ to $\hat R^2$ provide a negative weight. Above a threshold, the complex fixed points become real and one of them enters the physical region, where it is UV attractive in all directions and has $\alpha^*\neq0$.
\end{enumerate}

Compared with Riemannian quadratic gravity, the new element is the Weyl gauge boson. Its coupling does not spoil asymptotic freedom; on the contrary, it is asymptotically free for any two-derivative matter content. Its presence makes the fixed points of the Riemannian theory unreachable for $\alpha\neq0$ and creates a new attractive line, $r=(810+\Nc)/80$, on which the fate of the theory is decided by the $\hat R^2$ sector. The vanishing of the Weyl gauge coupling found in Ref.~\cite{Pagani:2013fca} for a theory linear in the curvature is replaced here by its logarithmic approach to zero.

Several directions remain open. The running of the Higgs couplings $\xi_h$ and $\lambda_H$, which requires a nonvanishing Higgs background, would allow one to study whether the electroweak scale generated in the broken phase~\cite{Ghilencea:2021lpa} is stable under the flow. The beta functions of the couplings of the four-derivative scalars are needed to establish the fixed point of Sec.~\ref{sec:HD} as a genuine UV completion. On the line $r=(810+\Nc)/80$ without additional matter, it would be interesting to investigate whether the flow can be continued through $1/\xi^2=0$, along the lines of Ref.~\cite{Salvio:2017qkx}. It would also be interesting to compare with the momentum running of Refs.~\cite{Buccio:2024hys,Buccio:2024omv}, with a background-independent heat-kernel computation, and with the geometric regularization of Refs.~\cite{Ghilencea:2023sti,Ghilencea:2024usf}, which does not affect the one-loop pole coefficients computed here. Finally, the anomaly structure of the theory with matter deserves further study, since dilatations act nontrivially on the metric; the consistency conditions for the anomaly of gauged Weyl symmetry have been analyzed in two dimensions in Ref.~\cite{Zanusso:2023vkn}.
In non-gauged conformal gravity, the BRST quantization of the conformal mode, the four-dimensional Liouville action obtained by integrating the trace anomaly, and the condition for the absence of BRST anomalies have been discussed in Ref.~\cite{Kawai:2025xbv}; it would be interesting to extend this analysis to the gauged case and to compare it with the arguments of Refs.~\cite{Ghilencea:2023sti, Ghilencea:2024usf}.

\begin{acknowledgments}
Authors thank R.~Percacci for useful discussions on the model and related topics.
LP thanks A.~Platania and the Niels Bohr Institute for hospitality during
the completion of this work.
The work of DB is supported by the European Research
Council (ERC) under the European Union’s Horizon
2020 research and innovation program Grant agreement No.~101170215 (ProbeQG).
The work of FC, LP and OZ is also supported by INFN under
the Iniziativa Specifica FLAG.
\end{acknowledgments}


\appendix
\section{Spin projections of the counterterms}
\label{app:projections}

In this appendix we derive Eq.~\eqref{eq:ansatz}. Consider the local functional
\begin{equation}
S_u=\int\dd^4x\sqrt g\left[u_C\,C^2+u_R\,\hat R^2+u_F\,\hat F^2\right]
\label{eq:Su}
\end{equation}
on the backgrounds of Sec.~\ref{sec:method}, $g_{\mu\nu}=\delta_{\mu\nu}+\bar h_{\mu\nu}$ and $\omega_\mu=\bar\omega_\mu$, and write its part quadratic in $\bar h_{\mu\nu}$ and $\bar\omega_\mu$ as $\frac12\int\bar\Phi(-p)\,\mathcal K(p)\,\bar\Phi(p)$, so that $\mathcal K$ plays the role of $\Gamma^{(2)}$ in Eq.~\eqref{eq:projections}. Since all curvatures vanish on flat space, only the squares of the linearized curvatures contribute. With $\partial_\mu\to ip_\mu$,
\begin{equation}
\begin{aligned}
R^{(1)}&=\partial_\mu\partial_\nu\bar h^{\mu\nu}-\Box\bar h\;\to\;p^2P^T_{\mu\nu}\bar h^{\mu\nu}\,,\\
\hat R^{(1)}&=R^{(1)}-6\,\partial_\mu\bar\omega^\mu\,,\\
\hat F^{(1)}_{\mu\nu}&=\partial_\mu\bar\omega_\nu-\partial_\nu\bar\omega_\mu\,,
\end{aligned}
\label{eq:linearized}
\end{equation}
while the linearized Weyl tensor depends only on the transverse-traceless part of $\bar h_{\mu\nu}$, for which $R^{(1)}_{\mu\nu}=-\frac12\Box\bar h_{\mu\nu}$ and $R^{(1)}=0$. The three terms of Eq.~\eqref{eq:Su} then give:
\begin{itemize}
\item $C^2$: using the Gauss--Bonnet identity, $\int C^2=\int\big(2R_{\mu\nu}^2-\frac23R^2\big)=\frac12\int\bar h_{\mu\nu}\Box^2\bar h^{\mu\nu}$ at second order;
\item $\hat R^2$: the square of $R^{(1)}$, the term $36(\partial_\mu\bar\omega^\mu)^2$, and a mixing $-12R^{(1)}\partial_\mu\bar\omega^\mu$ between $\bar h_{\mu\nu}$ and $\bar\omega_\mu$, which does not enter the projections~\eqref{eq:projections};
\item $\hat F^2$: integrating by parts, $2\,\bar\omega_\mu(-\Box\delta^{\mu\nu}+\partial^\mu\partial^\nu)\bar\omega_\nu$.
\end{itemize}
Using $P^T_{\mu\nu}P^T_{\rho\sigma}=3P^{(0,s)}_{\mu\nu,\rho\sigma}$, one finds
\begin{equation}
\begin{aligned}
\mathcal K_{\bar h\bar h}&=u_C\,p^4P^{(2)}+6u_R\,p^4P^{(0,s)}\,,\\
\mathcal K_{\bar\omega\bar\omega}&=4u_F\,p^2P^T+72u_R\,p^2P^L\,.
\end{aligned}
\label{eq:kernel}
\end{equation}
Each invariant contributes to definite channels: $C^2$ only to spin-$2$, $\hat F^2$ only to the transverse channel, and $\hat R^2$ to both the spin-$0$ and the longitudinal channels. Since the projectors are orthogonal, with $\mathrm{tr}\,P^{(2)}=5$, $\mathrm{tr}\,P^{(0,s)}=1$, $\mathrm{tr}\,P^T=3$ and $\mathrm{tr}\,P^L=1$, the projections~\eqref{eq:projections} are
\begin{align}
\Pi_2&=u_C\,p^4\,, & \Pi_0&=6u_R\,p^4\,,\nonumber\\
\Pi_T&=4u_F\,p^2\,, & \Pi_L&=72u_R\,p^2\,.
\label{eq:Piu}
\end{align}
In particular $\Pi_L=(12/p^2)\,\Pi_0$, which is the Ward identity~\eqref{eq:ward}. The divergence~\eqref{eq:Gammadiv} has the form~\eqref{eq:Su} with
\begin{equation}
u_C=-\frac{\beta_{1/\eta^2}}{2\epsilon}\,,\quad
u_R=\frac{\beta_{1/\xi^2}}{48\epsilon}\,,\quad
u_F=-\frac{\beta_{1/\alpha^2}}{8\epsilon}\,,
\end{equation}
and Eq.~\eqref{eq:Piu} reproduces Eq.~\eqref{eq:ansatz}. The Gauss--Bonnet density and total derivatives do not contribute to two-point functions on flat space.

\section{Relation to other parametrizations}
\label{app:conventions}

For the Reader who wishes to compare with other results in the literature on quadratic gravity, we collect here the relations with other common parametrizations. Writing the Euclidean action as $\int\sqrt g\,[\frac{1}{2\lambda}C^2+\frac{1}{\rho}\hat R^2+\frac{1}{\gamma^2}\hat F^2]$, as in our computer implementation, one has
\begin{equation}
\lambda=\frac{\eta^2}{2}\,,\qquad\rho=-24\,\xi^2\,,\qquad\gamma=2\alpha\,.
\end{equation}
The ratio $w$ of Eq.~\eqref{eq:rw} is $w=-3\lambda/\rho$, which coincides with the variable $\omega$ of Refs.~\cite{Codello:2006in,Percacci:2017fkn} at $\alpha=0$, and the rescaled time $\dd\bar\tau=\lambda\,\dd t$ used there is related to ours by $\bar\tau=8\pi^2\tau$. In the conventions of \emph{a}gravity~\cite{Salvio:2014soa}, $f_2^2=\eta^2/2$ and $f_0^2=4\xi^2$. Finally, in the formulation of Ref.~\cite{Ghilencea:2021lpa} the metric has charge $1$, the Weyl gauge coupling $\alpha_{\rm G}$ appears in the covariant derivative, and the Weyl tensor is that of the connection $\tilde\Gamma$. After combining the two sources of the kinetic term of the Weyl boson, its Lagrangian reads $-\frac{1}{4\gamma_{\rm G}^2}F_{\mu\nu}^2$ with $1/\gamma_{\rm G}^2=1+6\alpha_{\rm G}^2/\eta^2$. Rescaling the field to our charge normalization one finds $\alpha=\alpha_{\rm G}\gamma_{\rm G}/2$; with this identification the mass $m_\omega^2=\frac32\alpha_{\rm G}^2\gamma_{\rm G}^2M_p^2$ of Ref.~\cite{Ghilencea:2021lpa} coincides with Eq.~\eqref{eq:scales}.
%
\section{Computational details}
\label{app:computation}
The calculations of Secs.~\ref{sec:betas}--\ref{sec:HD} were performed in
\textsc{Mathematica} with \textsc{xAct}~\cite{xAct},
\textsc{xPert}~\cite{Brizuela:2008ra} and \textsc{xTras}~\cite{Nutma:2013zea}. The calculation was carried out
independently by two of the authors with two separate codes, and the results
were found to agree. In the final stages of this work, the authors used
Claude (Anthropic) to assist with debugging the code and checking the
manuscript; all results were derived and verified by the authors.
The list of diagrams and their weights was generated from the field content
and the vertex structure, as described in Sec.~\ref{sec:method}. The most
expensive step is the tensor reduction of the bubble diagrams, in particular
the one with two internal gravitons: its vertices carry up to four
derivatives and, for generic gauge parameters, its propagators contain all
the spin projectors of
Eqs.~\eqref{eq:projectionSpin1}--\eqref{eq:projections}. Since individual diagrams are gauge dependent, the parameters $a$, $b$, $p_1$ and $p_2$ are arbitrary and do not cancel
until the sum over diagrams in each channel is taken, which makes the
intermediate expressions considerably larger than the final results. For the
bubbles with two internal gravitons this growth was so severe that we set $\zeta=0$ from the start, which is allowed because these diagrams contain no vertices of order $1/\zeta$; in all other diagrams $\zeta$ was kept finite and the Landau limit was taken after the loop integration. Gauge independence of the final result was checked with respect to the remaining four parameters. Table~\ref{tab:timing}
summarizes the size and cost of the computation. Code is available upon request to the authors due to its size.
\begin{table}[htb!]
\caption{Approximate computing time of the one-loop two-point functions,
grouped by the fields running in the loop and summed over the four channels
of Eq.~\eqref{eq:ansatz}. The bubbles with two internal gravitons are
reported with four free gauge parameters (here abbreviated GP), as used for the final
results, and with one, for comparison; all other diagrams were computed with
all gauge parameters free. $\langle hv\rangle$ denotes the mixed propagator, which is of order $\zeta$ but
contributes in the Landau limit through vertices of order $1/\zeta$. Diagrams
that vanish identically are not counted. Times refer to a single core of an
Apple M3 Pro with 18 GB of RAM.}
\vspace{4pt}
\label{tab:timing}
\begin{ruledtabular}
\begin{tabular}{llcr}
Type & Loop fields & Diagrams & Time \\
\hline
Bubble (4 GP) & $hh$ & 2 & 110 h \\
Bubble (1 GP) & $hh$ & 2 & 6 h \\
Bubble & $hv$ & 2 & 18 h \\
Bubble & $\langle hv\rangle$, $vv$, $\varphi\varphi$ & 6 & 5 min \\
Bubble & ghosts ($c\bar c$, $\vartheta\bar\vartheta$, $c\bar\vartheta$) & 8 & 5 min \\
Bubble & third ghost $zz$ & 2 & 2 min \\
Tadpole & $hh$ & 2 & 1 h \\
Tadpole & all others & 4 & 1 min \\
\hline
Total (4 GP) & & 26 & $\sim 129$ h \\
Total (1 GP) & & 26 & $\sim 25$ h \\
\end{tabular}
\end{ruledtabular}
\end{table}
\pagebreak

\bibliographystyle{apsrev4-2}
\bibliography{biblio}

\end{document}